\pdfoutput=1
\documentclass[11pt,a4paper]{article}

\usepackage[T1]{fontenc}
\usepackage[utf8]{inputenc}
\usepackage{lmodern}
\usepackage{textcomp}
\usepackage{amsmath,amssymb}
\usepackage[a4paper,top=2.6cm,bottom=2.8cm,left=2.7cm,right=2.7cm,headsep=0.9cm]{geometry}
\usepackage{microtype}
\usepackage[english]{babel}

\usepackage{graphicx}
\usepackage{booktabs,tabularx,array,ragged2e}

\usepackage[font=small,labelfont=bf,labelsep=period,skip=6pt,justification=justified,singlelinecheck=false]{caption}
\usepackage{placeins}   
\usepackage{enumitem}
\setlist{itemsep=2pt,topsep=4pt,parsep=0pt,leftmargin=1.6em}
\usepackage{titlesec}
\titleformat{\section}{\normalfont\large\bfseries}{\thesection}{0.8em}{}
\titleformat{\subsection}{\normalfont\normalsize\bfseries}{\thesubsection}{0.8em}{}
\titlespacing*{\section}{0pt}{18pt plus 4pt minus 2pt}{8pt plus 2pt}
\titlespacing*{\subsection}{0pt}{12pt plus 3pt minus 2pt}{5pt plus 2pt}

\usepackage[round,authoryear]{natbib}
\usepackage{xcolor}
\definecolor{linkblue}{HTML}{1F4E79}
\usepackage[colorlinks=true,linkcolor=linkblue,citecolor=linkblue,urlcolor=linkblue,
            pdftitle={Who sees the sea? Sea-view potential by façade and floor from cadastral LoD1 models on the Costa del Sol},
            pdfauthor={Gabriel Rojo}]{hyperref}
\usepackage{xurl}
\usepackage{fancyhdr}
\fancypagestyle{plain}{\fancyhf{}\fancyfoot[C]{\small\thepage}}

\begin{document}

\thispagestyle{plain}
\begin{center}
{\fontsize{24}{29}\selectfont\bfseries Who sees the sea?\par}
\vspace{0.55em}
{\Large\setlength{\baselineskip}{1.25\baselineskip} Sea-view potential by façade and floor from cadastral LoD1 models\\ on the Costa del Sol\par}
\vspace{1.4em}
{\large Gabriel Rojo}\\[0.35em]
{\small \href{https://casas.es/}{CASAS.ES}\quad\textperiodcentered\quad\href{mailto:grojo@alumni.gsb.stanford.edu}{grojo@alumni.gsb.stanford.edu}}\\[0.9em]
{\small Working paper \textperiodcentered{} Version of 10 November 2026 \textperiodcentered{} CC BY 4.0}
\end{center}
\vspace{0.6em}

\begin{abstract}
\noindent Views are among the most valued and least measured attributes of a home. We present an open-data method that estimates the sea-view potential of every free façade of every residential building at three floors (ground, middle and top; other floors take the nearest of them), from cadastral footprints with floor counts and a national terrain model. Run for 158,060 buildings of 21 municipalities on Spain's Costa del Sol in under an hour on two cores, it finds that 52\,\% of buildings see some sea from their top floor and 36\,\% have an open view (20\textdegree{} or more). Traced over a LiDAR surface in a pilot, scores agree with the model's inside cadastral footprints (96.7\,\% of viewpoints in the same class at 20\textdegree{}), but structures the cadastre does not record lower the sea seen from the top floor by 6.1–9.0\textdegree{} on average. Against the photographs of 93 homes for sale, labelled blind by AI models of two developers with ties decided by the author, the score has an AUC of 0.80 (95\,\% CI 0.70–0.89; 0.81 against the first-round labels); at 20\textdegree{} it is sensitive (0.90) but not specific (0.45). Dwelling figures depend on how dwellings are allocated to façades and floors: 42\,\% of dwellings are on a floor with open-view potential under the main rule, and 12–45\,\% across the allocation, geometry and vegetation scenarios tested. As an application, higher-income census sections tend to have more modelled view potential within municipalities, with uncertain size and fragile inference (about two effective clusters); on terrain alone the difference largely disappears. Code and façade-level data reproduce the main results.

\vspace{0.6em}

\vspace{0.6em}
\noindent\textbf{Highlights}
\begin{itemize}\setlength{\itemsep}{1pt}
\item Sea-view potential by façade and floor for 158,060 buildings from open data
\item Close agreement with a LiDAR surface inside footprints; unmapped structures add bias
\item Photo validation: AUC 0.80, balanced accuracy 0.68 at 20\textdegree{}, two-developer labels
\item Dwelling shares are allocation-dependent outputs, reported as scenarios
\item Richer sections have more modelled potential; few-cluster inference is fragile
\end{itemize}

\end{abstract}
\vspace{0.8em}

\section{Introduction}

In a block of eleven floors 300 m from the beach in Marbella, the sea-facing façade sees no sea at all from floors 1 and 6; from floor 11 the sea fills 134\textdegree{} of the horizon (Figure 1). Whether a home sees the sea depends on its floor, on the terrain and on the buildings in front of it, and no public dataset records it.

Views matter for prices and for health. Since \citeauthor{rosen1974}'s (\citeyear{rosen1974}) formulation of hedonic prices, studies have priced views from lot-level comparisons to large samples with detailed view classes \citep{benson1998,bourassa2004}; views to water attract the largest premiums \citep{bond2002,conroy2011,hindsley2013}, and in tall buildings the view changes from floor to floor \citep{yu2007,jim2009,wong2011}. A window onto nature was associated with faster recovery after surgery \citep{ulrich1984}, and visible blue space, rather than proximity to it, with lower psychological distress \citep{nutsford2016} and less depression in older adults \citep{dempsey2018}. The European daylight standard EN 17037 includes a requirement for the view out \citep{cen2018}.

Measurement is the weak link. Hedonic studies often rely on appraiser judgments, listing descriptions or a single dummy variable \citep{bourassa2004}; GIS studies compute viewsheds on terrain models \citep{lake2000,sander2009} or, at best, on LiDAR surfaces and 3D city models \citep{hamilton2010,helbich2013,turan2021}. Recent large-scale work reads sea views from listing text with 3D simulation \citep{gillespie2024}, renders window views from photorealistic models \citep{li2022} or computes view metrics for a whole building stock from a commercial digital twin \citep{swietek2023}. Most countries have none of these inputs.

This paper describes a model that needs only data that national agencies already publish: building footprints with the number of floors, from the cadastre, and terrain models, from the national mapping agency. Its unit is the free façade at a given floor. It traces lines of sight from every free façade octant of every residential building at three floors, the ground floor, the middle floor and the top floor; the other floors take the value of the nearest traced floor. The façade-and-floor potential is the quantity the model measures and the one we test; building and dwelling figures are aggregations of it under stated rules, and dwelling figures in particular depend on how dwellings are allocated to façades and floors. We develop the model in a pilot in Marbella, test it against a LiDAR surface and against the photographs of homes for sale, and run it for 21 coastal municipalities of the Costa del Sol, a coast of tall seafront blocks and hillside houses where homes are sold and priced for their views \citep{castro2023}.

The contributions are four:

\begin{enumerate}

\item \textbf{An open-data method by façade and floor.} Only a cadastre with floor counts and a national terrain model are needed; the model builds its own LoD1 blocks and processes no point clouds.

\item \textbf{Explicit aggregation.} Building potential and dwelling shares are defined as aggregations of façade-and-floor scores, and every dwelling figure is reported under several allocation scenarios, as a conditional model output rather than a measured prevalence.

\item \textbf{Several references and an explicit sensitivity analysis.} Angles are compared with a trace over a LiDAR surface, heights with LiDAR, the tracer with a standard viewshed algorithm, and sea views with listing photographs, relabelled blind to the model by AI models of two developers, with ties decided by the author. Numerical choices are separated from structural assumptions, which are also combined and carried to dwellings and to income.

\item \textbf{Regional scale and a geometric decomposition.} The model runs for 158,060 buildings in under an hour on two cores, and a second run on terrain alone separates the sea that buildings hide from the sea that relief hides, which we apply, with inference designed for few clusters, to the relation between view potential and income.

\end{enumerate}

Sections 2 to 5 review related work and present the data, the method and the computation; Section 6 gives the results and the application to income, Section 7 the validations and robustness checks, and Section 8 the uses and limitations.

\section{Related work}

\subsection{Views in housing and health research}

Hedonic models price views as attributes of a property \citep{rosen1974}. Views were first coded as yes/no attributes and then graded by type, scope and quality \citep{benson1998,bourassa2004}, with the highest premiums for water \citep{bond2002}. GIS made views measurable for every property \citep{lake2000}, showed the bias of omitting them \citep{paterson2002}, and separated what is visible from what is merely nearby \citep{walls2015} and the scope of a viewshed \citep{hindsley2013}. In high-rise markets the view depends on floor and orientation \citep{yu2007,jim2009}, LiDAR surfaces improve view measures \citep{hamilton2010,helbich2013}, and segmented listing panoramas price what windows show \citep{peng2025}. Health research has moved from distance to blue space towards its visibility \citep{nutsford2016,dempsey2018,white2021}, with eye-level greenery and ocean visibility modelled from surfaces \citep{qiang2019,labib2021} and window-view distance related to mental health \citep{wei2026}.

\subsection{Computing views}

The viewshed is the basic GIS operation for visibility \citep{de2003}; its results depend on the algorithm \citep{fisher1993} and on surface errors \citep{fisher1994}. Cities add sharp near occluders \citep{bartie2010}; trees are neither opaque nor transparent \citep{murgoitio2014,vukomanovic2018}; long sight lines need curvature and refraction \citep{hirt2010}; and large terrains are handled with efficient sweeps \citep{haverkort2009}. Building science grades window views by content, access and clarity \citep{ko2022}, and city-scale window views have been rendered from photorealistic models \citep{li2022,li2025}. Where no 3D city model exists, LoD1 blocks can be generated from footprints and floor counts \citep{biljecki2015,biljecki2016,biljecki2017}; national models exist where LiDAR and footprints meet \citep{peters2022}, and global LoD1 heights have an error of about 4 m in Europe \citep{zhu2025}.

\subsection{The closest precedents}

Four studies come closest (Table 1). \citet{biljecki2017} showed that usable LoD1 models can be generated from footprints and floor counts without elevation data. \citet{swietek2023} computed view metrics for every building in Switzerland from a national 3D model and related them to income sorting across agglomerations. \citet{gillespie2024} simulated the breadth and depth of sea views for more than 100,000 Irish listings and priced them. \citet{yasumoto2011} computed viewsheds for about 5,000 homes in Kyoto and found better views in more affluent areas. This paper does not claim the first large-scale or distributional view measure. Its contribution is a combination these studies do not offer together: inputs limited to a cadastre with floor counts and a national terrain model; the façade and the floor as the unit, with explicit, tested rules for passing to buildings and dwellings; evaluation against a LiDAR surface, LiDAR heights, a standard viewshed algorithm and blinded labels of listing photographs; and a geometric decomposition applied at census-section level.

\begin{table}[!htbp]
\centering
\footnotesize\setlength{\tabcolsep}{4pt}
\caption{Large-scale view measures compared with this work.}\label{tab:1}
\begin{tabularx}{\linewidth}{>{\hsize=0.926\hsize\RaggedRight\arraybackslash}X>{\hsize=1.019\hsize\RaggedRight\arraybackslash}X>{\hsize=1.481\hsize\RaggedRight\arraybackslash}X>{\hsize=0.556\hsize\RaggedRight\arraybackslash}X>{\hsize=1.019\hsize\RaggedRight\arraybackslash}X}
\toprule
\textbf{Study} & \textbf{Area and scale} & \textbf{How views are measured} & \textbf{Unit} & \textbf{Needs} \\ \midrule
\citet{biljecki2017} & Generic method & LoD1 models generated from footprints and attributes, without elevation data & Building & Footprints with attributes \\
\citet{gillespie2024} & Ireland; more than 100,000 sale listings & 3D GIS simulation of sea-view breadth and depth; natural language processing of listing text & Listing & 3D model; listing text \\
\citet{swietek2023} & Switzerland; whole building stock & Building-level view metrics related to commune incomes & Building & National 3D model (commercial digital twin) \\
\citet{li2022,li2025} & Hong Kong; districts & Window views rendered from a photorealistic city model and segmented & Window & Photorealistic city model \\
\citet{peng2025} & Wuhan; 3,041 apartments & Segmentation of real-estate panoramas & Apartment & Listing panoramas \\
This work & Costa del Sol, Spain; 158,060 residential buildings & Lines of sight over cadastral buildings and national terrain models, out to 30 km & Façade and floor (building and dwelling by explicit rules) & Cadastre with floor counts; national terrain model \\
\bottomrule
\end{tabularx}
\end{table}

\section{Data}

The model itself uses only two national open datasets: cadastral buildings and the elevation models of the national mapping agency (Table 2). Everything else is used to test it. CASAS.ES, which displays the scores, describes how it processes these and other sources in its \href{https://casas.es/en/methodology}{methodology}.

\begin{table}[!htbp]
\centering
\footnotesize\setlength{\tabcolsep}{4pt}
\caption{Data sources.}\label{tab:2}
\begin{tabularx}{\linewidth}{>{\hsize=1.481\hsize\RaggedRight\arraybackslash}X>{\hsize=0.926\hsize\RaggedRight\arraybackslash}X>{\hsize=1.019\hsize\RaggedRight\arraybackslash}X>{\hsize=0.741\hsize\RaggedRight\arraybackslash}X>{\hsize=0.833\hsize\RaggedRight\arraybackslash}X}
\toprule
\textbf{Dataset} & \textbf{Provider} & \textbf{Resolution and date} & \textbf{Licence} & \textbf{Role} \\ \midrule
Buildings and building parts (INSPIRE BU): footprints, floors above ground per part, year built, use, condition & Dirección General del Catastro & Vector; datasets of 21 August 2026, downloaded 15 October 2026, for 21 municipalities of the Costa del Sol (Málaga province) & Catastro licence, attribution required & Model \\
MDT05 digital terrain model & Instituto Geográfico Nacional (IGN), derived from PNOA-LiDAR & 5 m grid; tiles read 15–18 October 2026 & CC BY 4.0-compatible (Orden FOM/\allowbreak{}2807/\allowbreak{}2015) & Model: terrain and sea up to 2 km \\
MDT25 digital terrain model & IGN & 25 m grid; 190 $\times$ 115 km around the coast & as above & Model: relief from 2 to 30 km \\
Normalised surface models of buildings and vegetation (mdsn\_e025, mdsn\_v025) & IGN, first PNOA-LiDAR coverage & 2.5 m & as above & LiDAR reference surface and heights (Sections 7.1 and 7.2); tree variants \\
Nomenclátor Geográfico Básico de España (NGBE) & IGN & Points; 5,387 names within the pilot region & as above & Names of dominant summits \\
Energy performance certificates with declared orientation & Registro de Certificados Energéticos de Andalucía, through the CASAS.ES API & 953 buildings in the pilot, 359 of them single-family & Public register & Orientation (supplementary material) \\
Homes for sale with an exact or street-level map position: type, floor, year built, advertised sea view, description and photographs & thinkSPAIN (thinkspain.com), read 15 October 2026 & 403 listings from Manilva to Nerja & Portal content; not redistributed & Validation of sea views (Section 7.3) \\
Atlas de Distribución de Renta de los Hogares (ADRH): median income per consumption unit, by census section & Instituto Nacional de Estadística (INE) & Census sections of Málaga province; income year 2023 & Free reuse with attribution & Application to income (Section 6.4) \\
Census section boundaries at 1 January 2024 & INE & Polygons; 1,085 sections in Málaga province & Free reuse with attribution & Linking buildings to income \\
Population and Housing Census 2021: dwellings by section, main and non-main & INE & Census sections (provisional section results) & Free reuse with attribution & Second homes, as a control (Section 6.4) \\
\bottomrule
\end{tabularx}
\end{table}

All services were read through their public endpoints: the INSPIRE download service of the Catastro, and the IGN Web Coverage Services for elevation (servicios.idee.es/wcs-inspire/mdt) and surface models (wcs-mds.idee.es/mds). The listings were read from the portal's public pages with a script. Their text and photographs were kept in a private archive and used only to match each home to a building and to label its photographs (Section 7.3); no listing text or photograph is redistributed, and the released validation tables identify neither listing nor building. The LiDAR layers are those of the first PNOA-LiDAR coverage, which the IGN serves ready-made; later coverages exist only as point clouds, which the method avoids processing.

\section{Method}

The model traces lines of sight from every free façade of every residential building, at three floors, over a surface of terrain plus cadastral buildings, and out to 30 km over the relief. It computes visibility where the observer is, at the façade rather than the parcel centroid \citep{yu2007,turan2021}, keeps near occluders sharp \citep{bartie2010} and keeps the far horizon, where mountains and the sea line lie \citep{qiang2019}.

\subsection{Surface model}

\begin{itemize}

\item \textbf{Terrain:} MDT05 is resampled bilinearly to a 2 m grid, giving the terrain elevation z\_T of each cell. MDT05 is derived from airborne LiDAR, but the method processes no point clouds and needs no surface model of buildings or trees.

\item \textbf{Buildings:} every cadastral building part with N $\geq$ 1 floors above ground is rasterised at 2 m as a solid block, with a height above terrain of h = 3.0 m $\times$ N + 1.0 m; the last metre stands for roofs and parapets. The surface is z\_T + h on the cells of a part and z\_T elsewhere, so roofs follow the terrain cell by cell. Where parts overlap, the tallest wins. The parts form an LoD1 block model of each building \citep{biljecki2016}, derived from floor counts as in \citet{biljecki2017}.

\item \textbf{Blocks:} each separate footprint of a cadastral building (the blocks of a complex on one parcel, a house and its pool) is a block, with its own façades and viewpoints. Building parts go to the block they overlap most, and blocks without a part with floors get no viewpoints. Parts recorded with more than 40 floors (one in the study area, with 100) are dropped as errors.

\item \textbf{Own block:} a ray treats the cells of the block it starts from as terrain along its whole length, so that the viewpoint, 0.75 m outside the wall, is not blocked by its own wall; other blocks of the same building are obstacles like any other. The cost is that another wing of a concave block does not block its own views; Section 7.4 tests the opposite extreme.

\item \textbf{Sea:} cells at 0.2 m or less that are connected to the edge of the grid, on MDT05 in the near field and on MDT25 in the far field, so both fields detect the sea with the same rule and no coastline vector is needed. Cells without elevation data are never sea.

\end{itemize}

\subsection{Free façades and viewpoints}

\begin{enumerate}

\item Walk the outer ring of each block of each residential building in use. An edge whose midpoint lies less than 0.30 m from another building or block is a party wall and is dropped.

\item Group the free edges by the octant of their outward normal (N, NE, E, SE, S, SW, W, NW). Octants with less than 3 m of free façade are dropped.

\item For each octant, place a viewpoint 0.75 m outside the middle of its longest edge, at floor 1, floor $\lceil$N/2$\rceil$ and floor N, where N is the floor count of the tallest part of the block within 6 m of the viewpoint, so that the one-floor terraces and porches drawn along many perimeters do not stand for the block behind them. Eye height = z\_T + 3.0 m $\times$ (floor \textminus{} 1) + 1.6 m, with z\_T the terrain of the viewpoint's cell.

\end{enumerate}

Only these three floors are traced. Every other floor takes the value of the nearest traced floor of the same façade (ties to the lower one); Section 4.4 gives the alternatives, and Section 7.3 traces the exact advertised floor of the photographed flats. Floors are numbered from 1, the ground floor, as in the cadastre's count of floors above ground. Spanish listings call the ground floor 0 (\emph{bajo}), so listing floor k is model floor k + 1.

\subsection{Lines of sight}

From each viewpoint, 86 rays leave at bearings 1\textdegree{}, 3\textdegree{}, …, 85\textdegree{} on either side of the façade normal. Each ray stands for the 2\textdegree{} sector centred on it, so together they cover the normal $\pm$ 86\textdegree{}, the half-plane in front of the façade of an observer at a balcony rail, and a score of \emph{s} degrees means that \emph{s}/2 rays reach the sea; scores run from 0\textdegree{} to 172\textdegree{}. This is much wider than the horizontal sight angles used to grade window views \citep{cen2018}; Section 7.4 keeps only the rays within $\pm$ 60\textdegree{} and $\pm$ 45\textdegree{} of the normal.

\begin{itemize}

\item \textbf{Near field, up to 2 km:} buildings and terrain on the 2 m grid, sampled every max(1 m, 0.004 $\times$ distance), that is every 8 m at 2 km. A ray that crosses less than one step of a building can miss it; this needs a narrow building, far away, crossed at a grazing angle, and halving the step changes the class of 0.3–0.4\,\% of buildings (Section 7.4).

\item \textbf{Far field, 2 to 30 km:} terrain only, on MDT25, sampled every 25 m from exactly 2 km. Both terrain models are IGN products in the same vertical datum; averaged over 25 m cells of the pilot, MDT25 lies a median 0.6 m above MDT05 (5th–95th percentiles \textminus{}3.7 to 5.0 m), which at 2 km changes an elevation angle by less than 0.15\textdegree{}, and moving the seam to 4 km changes the class of at most 1.2\,\% of buildings in any tile tested (Section 7.4).

\end{itemize}

At each sample at distance d, the tangent of the elevation angle is computed with Earth curvature and a refraction coefficient k = 0.13 \citep{hirt2010}, with z, d and z\_eye in metres and R = 6,371,000 m:

\begin{equation}
t(d) = \frac{z(d) - \dfrac{(1-k)\,d^{2}}{2R} - z_{\mathrm{eye}}}{d}
\end{equation}

The sea surface is placed at z = 0 exactly, not at the cell's terrain value (at most 0.2 m): a sea cell counts as visible when t(d) with z = 0 is higher than every tangent before it on the same ray, the far-field horizon included. Each ray stores the highest tangent within 300 m, within 2 km and beyond 2 km, the distance to the first visible sea and the distance of the highest far-field point. The tracer is written in Python with NumPy and SciPy and compiled with Numba \citep{harris2020,virtanen2020,lam2015}.

\begin{figure}[!htbp]
\centering
\includegraphics[width=0.95\linewidth]{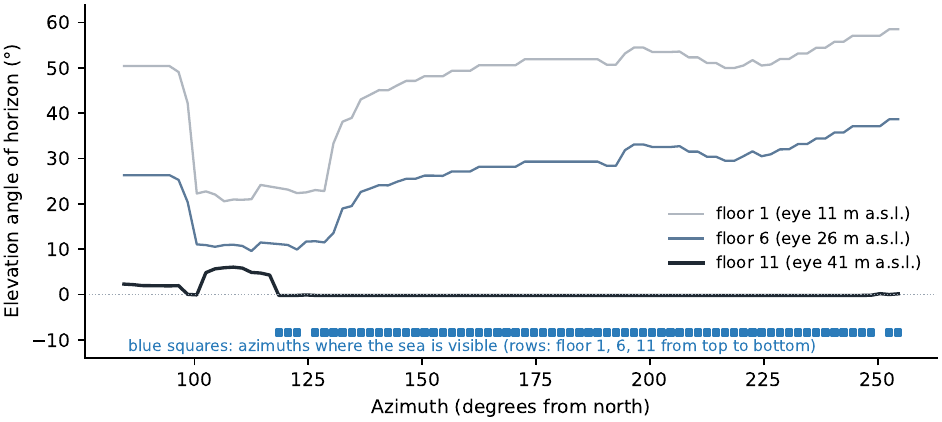}
\caption{How the model sees: horizon from the sea-facing façade of one 11-floor block at floors 1, 6 and 11 (elevation angle of the highest obstacle per bearing). Squares mark the bearings where the sea is visible. The first visible sea is 283 m away.}\label{fig:1}
\end{figure}

\subsection{Metrics and units}

Per viewpoint, the model gives the degrees of sea (2\textdegree{} per ray that reaches it), their mean bearing and the distance to the nearest visible sea. A distance-weighted variant counts each sea ray as min(1, 1 km / d), with d the distance to its first visible sea: seen from a given height, the sea surface fills an angle that falls roughly as 1/d, so the beach in front counts fully and a slot of sea 10 km away as a tenth. The same rays give the openness of the near field and the dominant mountains, released but not validated (supplementary material).

\textbf{Three units of measurement.} The model measures façade-and-floor potential; buildings and dwellings are aggregations of it. Let $a$ index the free-façade viewpoints of building $b$, with free length $L_a$, floor count $N_a$, traced floors $1$, $\lceil N_a/2 \rceil$ and $N_a$, and $s_a(f)$ degrees of sea at traced floor $f$.

\emph{Façade-and-floor potential} is $s_a(f)$ itself, the quantity compared with the LiDAR surface (Section 7.2) and, for the floor of each photographed dwelling, with the photographs (Section 7.3).

\emph{Building potential.} At level $\ell$ (ground, middle or top), each façade is read at its own floor for that level and the building takes the best façade:

\begin{equation}
S_b^{\ell} = \max_{a} \, s_a\left(f_{\ell}(a)\right), \qquad f_{\mathrm{ground}}(a) = 1, \quad f_{\mathrm{middle}}(a) = \lceil N_a/2 \rceil, \quad f_{\mathrm{top}}(a) = N_a
\end{equation}

A building sees the sea from a level when $S_b^{\ell} \geq 4^{\circ}$ (two rays), and has an open view when $S_b^{\ell} \geq 20^{\circ}$. These thresholds are conventions of the paper, not validated angles. All building figures use this definition.

\emph{Dwelling potential.} The cadastre gives the number of dwellings $D_b$ of a building but not their floors or façades, so dwellings need two assumptions. First, they are spread evenly over floors $1$ to $N_b = \max_a N_a$. Second, at floor $f$ the façades present are $A_b(f) = \{a : N_a \geq f\}$, and each takes the value $\hat{s}_a(f)$ of its nearest traced floor (the main rule), of the nearest traced floor below, or a linear interpolation between traced floors. For a threshold $\tau$, the share of the dwellings of floor $f$ with view potential is, under the best-façade and the length-share rules,

\begin{equation}
x_b^{\mathrm{best}}(f) = \mathbf{1}\left[ \max_{a \in A_b(f)} \hat{s}_a(f) \geq \tau \right], \qquad x_b^{\mathrm{share}}(f) = \frac{\sum_{a \in A_b(f)} L_a \, \mathbf{1}\left[ \hat{s}_a(f) \geq \tau \right]}{\sum_{a \in A_b(f)} L_a}
\end{equation}

and the share of dwellings with view potential in a set $G$ of buildings is

\begin{equation}
P_G = \frac{\sum_{b \in G} \left( D_b / N_b \right) \sum_{f=1}^{N_b} x_b(f)}{\sum_{b \in G} D_b}
\end{equation}

The best-façade rule gives each dwelling the best free façade of its floor, which is optimistic for a single dwelling; the length-share rule assumes that a floor's dwellings face its free façades in proportion to their length, which is more conservative but not a lower bound, because a dwelling with windows on two façades sees the sea from either. The rules are allocation scenarios, not a confidence interval, and every dwelling figure is a model output conditional on them and on the geometry (Section 7.4).

\textbf{Joining dwellings to floors.} The even spread over floors is the weakest assumption of Eq. (4), and the cadastre can replace it. Its alphanumeric records of non-protected data (CAT format) list every property of a parcel, with a 20-character reference whose first 14 characters are the parcel reference that identifies the building in the INSPIRE data, and the floor and door of each unit (records of type 14 and 15). The released façade table carries, for every viewpoint, that 14-character reference, the octant, the free length and floor count of the façade and the traced floor. Counting each building's dwellings by floor in the CAT records and using those counts in place of $D_b / N_b$ would remove the dwellings that Eq. (4) puts in ground-floor shops and weight penthouses correctly; matching doors to façades would still need floor plans. The CAT files are not part of the INSPIRE download used here, and the join is left to users of the data.

\subsection{Parameters}

\begin{table}[!htbp]
\centering
\small
\caption{Model parameters.}\label{tab:3}
\begin{tabularx}{\linewidth}{>{\hsize=1.130\hsize\RaggedRight\arraybackslash}X>{\hsize=0.870\hsize\RaggedRight\arraybackslash}X}
\toprule
\textbf{Parameter} & \textbf{Value} \\ \midrule
Grid of the near field & 2 m \\
Height of a building part & 3.0 m $\times$ floors + 1.0 m, above the terrain of each cell \\
Own block & transparent along the whole ray; other blocks of the building are obstacles \\
Floors of a façade & tallest part of the block within 6 m of the viewpoint \\
Traced floors & 1, $\lceil$N/\allowbreak{}2$\rceil$ and N; other floors from the nearest traced floor \\
Eye height & terrain at the viewpoint cell + 3.0 m $\times$ (floor \textminus{} 1) + 1.6 m \\
Viewpoint offset from the façade & 0.75 m \\
Party-wall tolerance & 0.30 m \\
Minimum free façade per octant & 3 m \\
Rays & 86, at $\pm$ 1\textdegree{}, $\pm$ 3\textdegree{}, …, $\pm$ 85\textdegree{} from the normal; each counts 2\textdegree{} (scores 0–172\textdegree{}) \\
Near field (buildings and MDT05); step & 2 km; max(1 m, 0.004 d) \\
Far field; step & 2–30 km; 25 m (MDT25) \\
Sea cells & $\leq$ 0.2 m, connected to the open sea; sea surface at z = 0 \\
Refraction coefficient & 0.13 \\
\bottomrule
\end{tabularx}
\end{table}

\section{Study areas and computation}

\textbf{Pilot.} The pilot covers 6 $\times$ 5.5 km of Marbella (UTM zone 30N, 328,000–334,000 E $\times$ 4,040,500–4,046,000 N), from the old town and the seafront blocks to the hillside villas of Nagüeles and Sierra Blanca, from sea level to about 400 m, under La Concha (1,215 m). Its near-field surface covers 10 $\times$ 10 km at 2 m and its far field 70 $\times$ 70 km of MDT25. It has 5,231 residential buildings (463 of six floors or more), 26,178 free façade octants and 52,971 viewpoints, traced in 71 s on two processor cores.

\textbf{Whole coast.} The default model (terrain and buildings, without trees) was run for every residential building in use in 21 municipalities, from Manilva to Nerja, including the city of Málaga and two inland towns, Alhaurín de la Torre and Alhaurín el Grande: 159,152 residential buildings in the cadastre, in 159 tiles of 4 km, each with a 2 km margin of neighbouring buildings and terrain. 158,060 buildings (99.3\,\%) received viewpoints, 1.35 million in all; the others have no block with 3 m of free façade in one octant, or no floor above ground. Ray tracing took 2,504 s and data preparation 237 s on two cores (two virtual cores of an Intel Xeon at 2.1 GHz, 8 GB of memory; Python 3.11, NumPy 2.4, Numba 0.68), or 9.6 core-hours per million buildings. The time is per tile, from assembling its buildings to the building summary; it excludes the one-off download and conversion of the inputs and the dwelling, income and validation analyses.

\textbf{Counterfactual without buildings.} To measure how much sea the buildings hide, the whole coast was run again with the same façades, floors and viewpoints over terrain alone, so that no building, not even another part of the same building, blocks a line of sight.

\textbf{Variants.} For Section 7.4, the whole coast was also run with LiDAR trees, opaque and at half their height; with the floors of a façade taken within 1.5 m; and with two combinations of conservative assumptions (1.5 m radius, own block opaque beyond 4 m and flat roofs, once with trees at half height and once with opaque trees). Narrower fields of view and the distance-weighted extent are read from the same rays. The pilot was also traced over a LiDAR surface (Section 7.2), and the 92 buildings of the photographed sample were traced again to add the advertised floor of each flat and the variants that the coast runs do not cover.

\section{Results}

\subsection{Sea views depend on the floor and on the terrain}

With buildings and terrain, 57\,\% of the pilot's residential buildings see some sea from the ground floor, 64\,\% from the middle floor and 78\,\% from the top floor; an open sea view (20\textdegree{} or more) goes from 36\,\% at ground level to 62\,\% at the top. When a building sees the sea from its top floor, the nearest visible sea is a median 1.7 km away. These are building potentials (Eq. 2): at least one free façade at that level has the view, which is generous for any single dwelling (Section 7.3).

\begin{figure}[!htbp]
\centering
\includegraphics[width=1.00\linewidth]{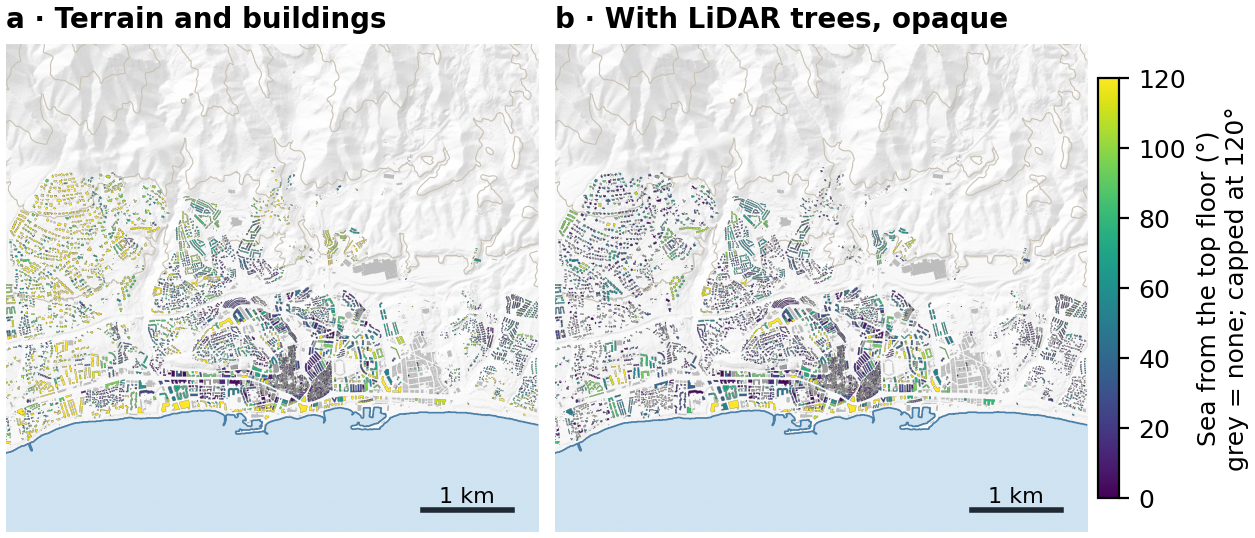}
\caption{Degrees of sea visible from the top floor of each residential building in the pilot. a: buildings and terrain. b: adding LiDAR trees as opaque obstacles (Section 6.2). Grey: no sea. Background: hillshade of MDT05, contours at 100 m and every 200 m from 200 m, and the coastline of the sea mask (blue line).}\label{fig:2}
\end{figure}

The floor matters most in the dense town. In buildings of six floors or more (n = 463), 38\,\% see the sea from the ground floor, 66\,\% from the middle floor and 93\,\% from the top (Figure 3a); for an open view, 15\,\%, 34\,\% and 77\,\%. The block of Figure 1 sees no sea from floors 1 and 6, and 134\textdegree{} of sea from floor 11. This is the vertical dimension that hedonic studies of high-rise markets price \citep{jim2009,wong2011} and that a building-level score cannot express. The terrain matters most outside the town (Table 4): the low ground 10 to 25 m above sea level, behind the seafront blocks, is where views are rarest, and above 100 m almost every building sees the sea, as \citet{castro2023} observed for Marbella's slopes. The same rays describe the rest of the horizon: from the top floor, 90\,\% of the pilot's buildings see relief beyond 2 km rising above everything nearer, most often La Concha (supplementary Section S1).

\begin{table}[!htbp]
\centering
\small
\caption{Share of buildings that see the sea, by the elevation of the ground they stand on (pilot).}\label{tab:4}
\begin{tabularx}{\linewidth}{>{\hsize=1.042\hsize\RaggedRight\arraybackslash}X>{\hsize=0.833\hsize\raggedleft\arraybackslash}X>{\hsize=0.938\hsize\raggedleft\arraybackslash}X>{\hsize=0.938\hsize\raggedleft\arraybackslash}X>{\hsize=1.250\hsize\raggedleft\arraybackslash}X}
\toprule
\textbf{Ground elevation (m)} & \textbf{Buildings} & \textbf{Ground floor} & \textbf{Top floor} & \textbf{Median sea from top floor (\textdegree{})} \\ \midrule
0–10 & 162 & \mbox{72\,\%} & \mbox{86\,\%} & 55 \\
10–25 & 1,015 & \mbox{26\,\%} & \mbox{46\,\%} & 2 \\
25–50 & 1,410 & \mbox{41\,\%} & \mbox{70\,\%} & 16 \\
50–100 & 1,760 & \mbox{67\,\%} & \mbox{90\,\%} & 44 \\
100–200 & 787 & \mbox{94\,\%} & > \mbox{99\,\%} & 98 \\
200–600 & 97 & \mbox{100\,\%} & \mbox{100\,\%} & 128 \\
\bottomrule
\end{tabularx}
\end{table}

\begin{figure}[!htbp]
\centering
\includegraphics[width=1.00\linewidth]{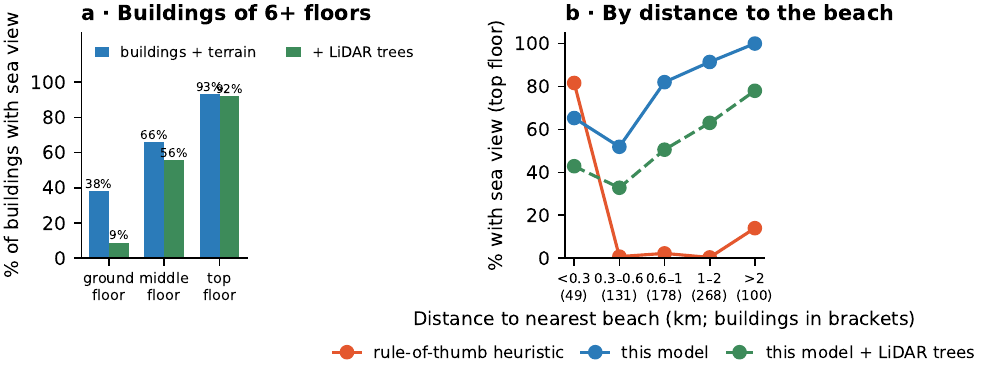}
\caption{a: share of buildings of six floors or more that see the sea, by floor, with and without LiDAR trees. b: share of the 726 sampled buildings with a sea view from the top floor, by distance to the nearest beach, for a rule-of-thumb heuristic and the model with and without trees (supplementary Section S3).}\label{fig:3}
\end{figure}

\subsection{Trees matter more for houses than for flats}

Trees are the largest structural uncertainty. With LiDAR vegetation as opaque obstacles, the share of the pilot's buildings that see the sea falls from 57\,\% to 12\,\% at ground level and from 78\,\% to 55\,\% at the top floor; with trees at half their height, to 35\,\% and 74\,\%. Tall buildings keep their top-floor views (93\,\% against 92\,\%) but lose them lower down (Figure 3a). The default model leaves trees out, because they are partly transparent \citep{murgoitio2014}, gardens are pruned to keep views and the vegetation layer is about a decade old, yet trees do hide views in low-density settings \citep{vukomanovic2018}. The tree variants are reported beside the default throughout (Sections 7.3 and 7.4); against the photographs, trees at half height rank the homes slightly better than the default (Section 7.3).

\subsection{The whole Costa del Sol}

Across the coast, 38\,\% of residential buildings see some sea from the ground floor and 52\,\% from the top floor; 24\,\% and 36\,\% have an open view (Figure 4). Weighted by their dwellings, 59\,\% of the 810,334 dwellings are in buildings with open-view potential at the top floor. The share of dwellings on a floor with open-view potential, which needs the allocation of Eqs. (3) and (4), is 42\,\% under the main rule and 21\,\% under the length-share rule (Section 6.4). From the top floor, the nearest visible sea is a median 1.6 km away for buildings with an open view and 4.4 km for those that see less than 20\textdegree{}.

\begin{figure}[!htbp]
\centering
\includegraphics[width=1.00\linewidth]{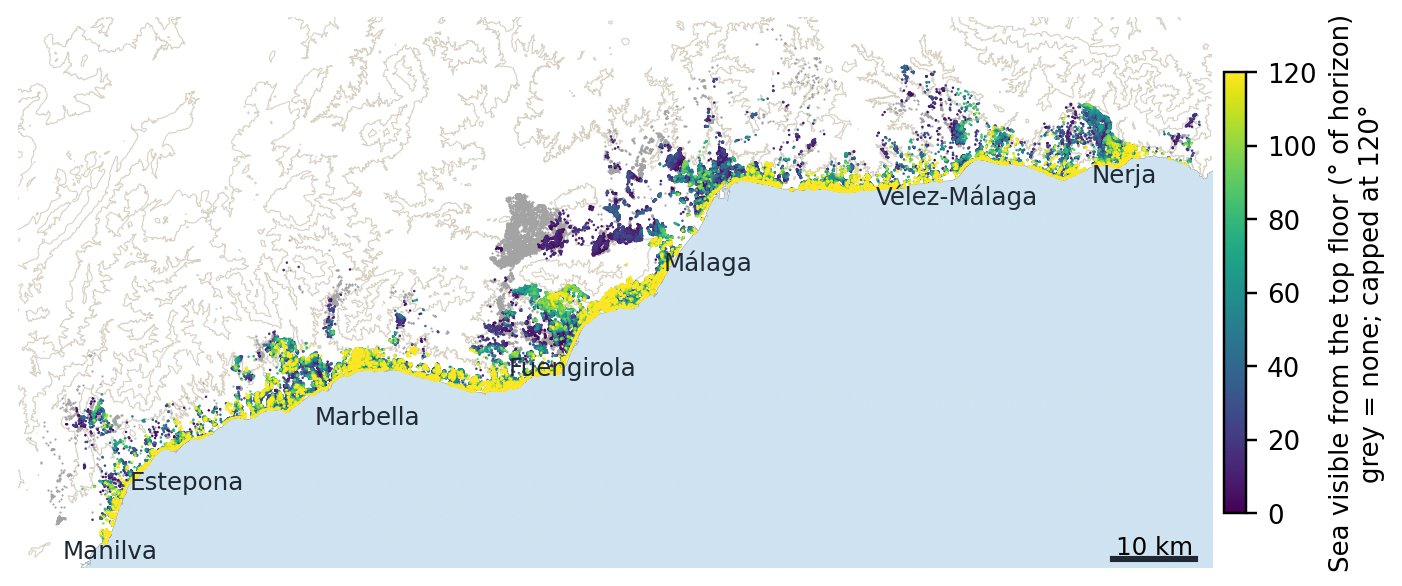}
\caption{Degrees of sea visible from the top floor of each of the 158,060 residential buildings of 21 municipalities of the \href{https://casas.es/en/malaga/costa-del-sol-comarca/}{Costa del Sol} (best façade; terrain and buildings, no trees). Grey: no sea. Contours every 400 m from 200 m.}\label{fig:4}
\end{figure}

The coast is far from uniform (Table 5). In Benahavís, Benalmádena, Rincón de la Victoria and Manilva, where hills rise from the shore, three buildings in five or more have an open view from the top floor; in Málaga, where most buildings stand in the flat city behind the port, 18\,\% do, and in Fuengirola, a dense town on a narrow plain, 22\,\%. Distance to the sea explains less than one might expect: buildings within 500 m of the sea have an open top-floor view in 52\,\% of cases, and those 2 to 5 km away in 35\,\%, mostly on slopes facing the sea. In the 1,760 buildings of ten floors or more, 14\,\% have an open view from the ground floor and 73\,\% from the top.

\begin{table}[!htbp]
\centering
\footnotesize\setlength{\tabcolsep}{4pt}
\caption{Share of residential buildings that see the sea, by municipality, from west to east. The last column is the counterfactual of Section 6.5, with terrain only. Each municipality links to its page at casas.es.}\label{tab:5}
\begin{tabularx}{\linewidth}{>{\hsize=1.474\hsize\RaggedRight\arraybackslash}X>{\hsize=0.789\hsize\raggedleft\arraybackslash}X>{\hsize=0.842\hsize\raggedleft\arraybackslash}X>{\hsize=0.842\hsize\raggedleft\arraybackslash}X>{\hsize=0.842\hsize\raggedleft\arraybackslash}X>{\hsize=1.000\hsize\raggedleft\arraybackslash}X>{\hsize=1.211\hsize\raggedleft\arraybackslash}X}
\toprule
\textbf{Municipality} & \textbf{Buildings} & \textbf{Dwellings} & \textbf{Any sea, ground floor} & \textbf{Any sea, top floor} & \textbf{Open view ($\geq$ 20\textdegree{}), top floor} & \textbf{Open view ($\geq$ 20\textdegree{}), top floor, without buildings} \\ \midrule
\href{https://casas.es/en/malaga/casares/}{Casares} & 2,249 & 7,523 & \mbox{31\,\%} & \mbox{40\,\%} & \mbox{28\,\%} & \mbox{40\,\%} \\
\href{https://casas.es/en/malaga/manilva/}{Manilva} & 2,981 & 17,911 & \mbox{57\,\%} & \mbox{76\,\%} & \mbox{60\,\%} & \mbox{92\,\%} \\
\href{https://casas.es/en/malaga/estepona/}{Estepona} & 8,390 & 52,044 & \mbox{47\,\%} & \mbox{63\,\%} & \mbox{47\,\%} & \mbox{94\,\%} \\
\href{https://casas.es/en/malaga/benahavis/}{Benahavís} & 2,388 & 9,877 & \mbox{79\,\%} & \mbox{82\,\%} & \mbox{72\,\%} & \mbox{79\,\%} \\
\href{https://casas.es/en/malaga/istan/}{Istán} & 904 & 1,254 & \mbox{23\,\%} & \mbox{24\,\%} & \mbox{19\,\%} & \mbox{20\,\%} \\
\href{https://casas.es/en/malaga/marbella/}{Marbella} & 16,765 & 100,176 & \mbox{59\,\%} & \mbox{75\,\%} & \mbox{57\,\%} & \mbox{95\,\%} \\
\href{https://casas.es/en/malaga/ojen/}{Ojén} & 1,146 & 3,524 & \mbox{32\,\%} & \mbox{54\,\%} & \mbox{28\,\%} & \mbox{44\,\%} \\
\href{https://casas.es/en/malaga/alhaurin-el-grande/}{Alhaurín el Grande} & 9,104 & 12,698 & \mbox{3\,\%} & \mbox{4\,\%} & \mbox{0\,\%} & \mbox{0\,\%} \\
\href{https://casas.es/en/malaga/mijas/}{Mijas} & 16,743 & 63,934 & \mbox{54\,\%} & \mbox{64\,\%} & \mbox{48\,\%} & \mbox{81\,\%} \\
\href{https://casas.es/en/malaga/fuengirola/}{Fuengirola} & 5,713 & 50,935 & \mbox{23\,\%} & \mbox{34\,\%} & \mbox{22\,\%} & \mbox{96\,\%} \\
\href{https://casas.es/en/malaga/alhaurin-de-la-torre/}{Alhaurín de la Torre} & 8,443 & 16,129 & \mbox{22\,\%} & \mbox{36\,\%} & \mbox{9\,\%} & \mbox{35\,\%} \\
\href{https://casas.es/en/malaga/benalmadena/}{Benalmádena} & 6,670 & 49,512 & \mbox{66\,\%} & \mbox{79\,\%} & \mbox{63\,\%} & \mbox{97\,\%} \\
\href{https://casas.es/en/malaga/torremolinos/}{Torremolinos} & 3,710 & 44,640 & \mbox{46\,\%} & \mbox{68\,\%} & \mbox{49\,\%} & \mbox{99\,\%} \\
\href{https://casas.es/en/malaga/malaga/}{Málaga} & 40,147 & 259,678 & \mbox{17\,\%} & \mbox{32\,\%} & \mbox{18\,\%} & \mbox{67\,\%} \\
\href{https://casas.es/en/malaga/totalan/}{Totalán} & 324 & 344 & \mbox{20\,\%} & \mbox{31\,\%} & \mbox{1\,\%} & \mbox{1\,\%} \\
\href{https://casas.es/en/malaga/rincon-de-la-victoria/}{Rincón de la Victoria} & 5,094 & 27,468 & \mbox{68\,\%} & \mbox{81\,\%} & \mbox{63\,\%} & \mbox{96\,\%} \\
\href{https://casas.es/en/malaga/velez-malaga/}{Vélez-Málaga} & 12,677 & 48,478 & \mbox{42\,\%} & \mbox{55\,\%} & \mbox{40\,\%} & \mbox{75\,\%} \\
\href{https://casas.es/en/malaga/algarrobo/}{Algarrobo} & 1,545 & 5,539 & \mbox{41\,\%} & \mbox{53\,\%} & \mbox{40\,\%} & \mbox{62\,\%} \\
\href{https://casas.es/en/malaga/torrox/}{Torrox} & 4,225 & 19,199 & \mbox{48\,\%} & \mbox{66\,\%} & \mbox{47\,\%} & \mbox{77\,\%} \\
\href{https://casas.es/en/malaga/frigiliana/}{Frigiliana} & 2,264 & 2,959 & \mbox{62\,\%} & \mbox{72\,\%} & \mbox{56\,\%} & \mbox{75\,\%} \\
\href{https://casas.es/en/malaga/nerja/}{Nerja} & 6,578 & 16,512 & \mbox{54\,\%} & \mbox{74\,\%} & \mbox{57\,\%} & \mbox{94\,\%} \\
\textbf{Costa del Sol} & \textbf{158,060} & \textbf{810,334} & \textbf{\mbox{38\,\%}} & \textbf{\mbox{52\,\%}} & \textbf{\mbox{36\,\%}} & \textbf{\mbox{73\,\%}} \\
\bottomrule
\end{tabularx}
\end{table}

\subsection{An application: view potential and income}

This section is an application of the model, not its validation. Each residential building was assigned to the census section that contains it, and each section to its median income per consumption unit in 2023 (Table 2); 155,366 buildings with 810,168 dwellings in 871 sections have both a view and an income.

In the poorest fifth of dwellings, ranked by the income of their section, 32\,\% are on a floor with open-view potential under the main allocation; in the richest fifth, 48\,\% (Figure 5). The poorest fifth does not live further from the sea (median distance 966 m against 1,025 m), and the two fifths have similar shares of single-family houses (16\,\% and 15\,\%) and heights (6.0 and 5.8 floors). The gap is largest 1 to 3 km from the sea (16\,\% against 51\,\%); within 1 km it is 48\,\% against 54\,\%.

\begin{figure}[!htbp]
\centering
\includegraphics[width=0.85\linewidth]{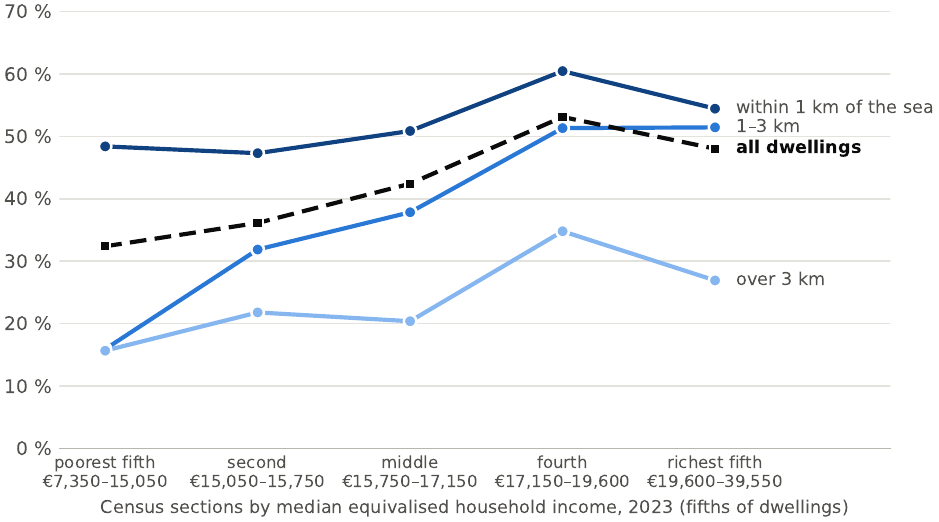}
\caption{Share of dwellings with open sea-view potential (20\textdegree{} or more) on their floor, under the main allocation (best façade, nearest traced floor; Eqs. 3 and 4), by fifth of dwellings ranked by the median income per consumption unit of their census section (INE income atlas for 2023), for all dwellings and by distance to the sea. Costa del Sol, 810,168 dwellings in 871 sections.}\label{fig:5}
\end{figure}

\textbf{Allocation scenarios.} Similar shares of flats and heights do not guarantee that the best-façade rule overstates rich and poor areas alike, so Table 6 repeats the analysis under each allocation scenario. Sharing each floor's dwellings among its façades lowers the share with open-view potential from 42\,\% to 21\,\%, from 32\,\% to 18\,\% in the poorest fifth and from 48\,\% to 24\,\% in the richest: the gap narrows from 16 to 6 points but does not close.

\begin{table}[!htbp]
\centering
\footnotesize\setlength{\tabcolsep}{4pt}
\caption{Share of dwellings with open-view potential (20\textdegree{} or more) under each allocation scenario (Eqs. 3 and 4), for all dwellings and for the poorest and richest fifths by section income, and the section-level income coefficient with municipality fixed effects (model M2 of Table A1; percentage points per unit of log income), with buildings and on terrain alone. In brackets, the 95\,\% interval from the restricted wild cluster bootstrap-t (WCR; 9,999 replications, Webb weights, clustered by municipality, interval by test inversion).}\label{tab:6}
\begin{tabularx}{\linewidth}{>{\hsize=1.860\hsize\RaggedRight\arraybackslash}X>{\hsize=0.783\hsize\raggedleft\arraybackslash}X>{\hsize=0.832\hsize\raggedleft\arraybackslash}X>{\hsize=0.734\hsize\raggedleft\arraybackslash}X>{\hsize=0.979\hsize\raggedleft\arraybackslash}X>{\hsize=0.783\hsize\raggedleft\arraybackslash}X>{\hsize=1.028\hsize\raggedleft\arraybackslash}X}
\toprule
\textbf{Allocation: façade / floor} & \textbf{All} & \textbf{Poorest fifth} & \textbf{Richest fifth} & \textbf{Income coefficient [WCR]} & \textbf{Terrain only: all} & \textbf{Terrain only: income coefficient [WCR]} \\ \midrule
Best / nearest (main) & \mbox{42.4\,\%} & \mbox{32.4\,\%} & \mbox{48.0\,\%} & 52.5 [14.4, 92.9] & \mbox{87.2\,\%} & 3.2 [\textminus{}18.6, 34.9] \\
Best / floor below & \mbox{39.9\,\%} & \mbox{29.5\,\%} & \mbox{45.1\,\%} & 54.3 [9.0, 97.9] & \mbox{86.8\,\%} & 3.9 [\textminus{}17.6, 35.0] \\
Best / interpolated & \mbox{44.0\,\%} & \mbox{33.9\,\%} & \mbox{49.4\,\%} & 52.6 [18.8, 90.6] & \mbox{87.6\,\%} & 2.6 [\textminus{}18.4, 32.6] \\
Length share / nearest & \mbox{21.2\,\%} & \mbox{18.3\,\%} & \mbox{23.8\,\%} & 23.2 [8.6, 38.2] & \mbox{59.4\,\%} & 0.3 [\textminus{}3.2, 16.4] \\
Length share / floor below & \mbox{19.4\,\%} & \mbox{16.1\,\%} & \mbox{21.8\,\%} & 24.1 [5.3, 41.0] & \mbox{58.8\,\%} & 1.0 [\textminus{}10.4, 17.2] \\
Length share / interpolated & \mbox{22.2\,\%} & \mbox{19.3\,\%} & \mbox{24.8\,\%} & 23.7 [8.9, 38.9] & \mbox{60.0\,\%} & \textminus{}0.3 [\textminus{}4.0, 14.6] \\
Best / nearest, no dwellings on the ground floor of buildings of 3+ floors & \mbox{44.9\,\%} & \mbox{34.5\,\%} & \mbox{50.7\,\%} & 54.0 [18.8, 91.8] & — & — \\
\bottomrule
\end{tabularx}
\end{table}

\textbf{Regressions.} One row per census section with at least 20 dwellings (860 sections in 21 municipalities), weighted by dwellings, explains the share of dwellings with open-view potential, in percent, by the log of section income and the log of the mean distance of its dwellings to the sea (Appendix A). The models answer different questions. M1 gives the association across the coast; M2 adds municipality fixed effects, which absorb planning regimes and the shape of each coast; M5 adds building age, type, height and ground elevation, which are partly the channels through which income and views go together rather than confounders. M2 and M5 are different conditional associations, not bounds on one quantity.

\textbf{Inference with few, unequal clusters.} Málaga holds 51\,\% of the sections, and its partial leverage for the income coefficient of M2 is 0.66, so the effective number of clusters \citep{carter2017} is about 2.2 out of 21 (supplementary Table S6). With so few effective clusters no cluster-robust method is reliable, and the usual ones fail in known directions \citep{mackinnon2023a,mackinnon2023b}: the restricted wild cluster bootstrap-t, WCR \citep{mackinnon2017}, tends to be conservative when one cluster dominates, its unrestricted version, WCU, anti-conservative, and leave-one-cluster-out errors, CV3, reflect how far the estimate moves when each municipality is dropped. We use WCR for every central comparison (Tables 6 and A3, and the scenarios of Section 7.4), report CV3 for the models of Table A1, add spatial errors that do not depend on municipal boundaries, and make no firm significance claim where defensible methods disagree.

\begin{itemize}

\item \emph{Within municipalities (M2).} 52.5 points per unit of log income (WCR 95\,\% interval 14.4 to 92.9, p = 0.026; CV3 with t(20), 9.1 to 95.8). A section with twice the income has about 36 points more dwellings with open-view potential at the same distance from the sea. Dropping any single municipality leaves it between 50.7 and 73.5; without Málaga (M7), the effective number of clusters rises to 7.0, and the coefficient, 73.5, has a WCR interval of 55.0 to 106.4. Errors that allow correlation between nearby sections across municipal boundaries \citep{conley1999}, with a Bartlett kernel, are 7.5, 9.3 and 8.7 for cut-offs of 2, 5 and 10 km, against 7.4 clustered by municipality, and bootstraps of square spatial blocks of 2.5, 5 and 10 km, each with four grid origins, give 95\,\% intervals whose lower bounds range from 35.2 to 42.5 and upper bounds from 67.7 to 91.2; none includes zero. The residuals keep some spatial correlation (Moran's I 0.27 between sections within 2 km).

\item \emph{Within Málaga.} The city alone has 435 sections, so this comparison needs no municipal clustering: 43.8 (95\,\% interval from a bootstrap of 2 km spatial blocks, 28.7 to 60.5; 36 blocks), and 19.7 (HC1 standard error 4.2) with the building controls of M5.

\item \emph{Across the coast (M1).} 30.2 (WCR \textminus{}1.8 to 85.4, p = 0.054).

\item \emph{With building controls (M5).} 16.5, and here the methods disagree: CV3 gives 7.1 to 26.0, WCR \textminus{}6.2 to 37.0 (p = 0.098) and WCU 10.1 to 23.0. The disagreement follows from the design, not from an error: dropping any single municipality moves the estimate only between 14.6 and 18.8, so the jackknife variance behind CV3 is small, while WCR, with Málaga carrying 0.65 of the partial leverage (2.3 effective clusters), is wide. We describe the M5 association as positive and stable to dropping any municipality, not as established.

\item \emph{Second homes.} The 2021 census share of non-main dwellings leaves M2 at 47.6 (M4).

\end{itemize}

\textbf{Outcomes and functional form.} With the best façade, the association excludes zero under WCR for any sea (4\textdegree{}: 62.5, WCR 45.3 to 92.5) and for open views, but not for wide views of 45\textdegree{} or more (37.1, WCR \textminus{}4.6 to 76.2) or for the mean degrees of sea (39.4, WCR \textminus{}2.8 to 82.8); under the length-share rule it excludes zero for every outcome (Table A3). The outcome is a share, bounded at 0 and 100\,\%; a fractional logit \citep{papke1996} with municipality fixed effects gives an average marginal effect of 58.7 points per unit of log income (CV3 24.4 to 92.9), against 52.5 in the linear model.

\textbf{Representation.} The size of the association depends on the allocation rule, trees, geometry and the weight given to distant sea (Section 7.4). With the best façade at 20\textdegree{} its point estimate goes from 28.0 to 52.5 across the coast-wide geometric and vegetation scenarios, and with the length-share rule from 9.7 to 23.2; its WCR interval excludes zero in every scenario except the distance-weighted extent, under both rules.

\textbf{What the association means.} The supported conclusion is narrow: higher-income census sections tend to have more modelled sea-view potential, within municipalities and at equal distance from the sea, with substantial uncertainty about its size and a clear dependence on representation. It is an association between areas, not households \citep{banzhaf2019}, consistent with amenity-based theories of location \citep{brueckner1999}. Resident income is not the income of all owners of the stock: second homes count as dwellings but their owners' incomes are not in the atlas, and residents who file taxes abroad, common among the coast's foreign residents, are missing from it. Income refers to 2023 and buildings to the cadastre of 2026, although only 1.3\,\% of dwellings are in buildings completed in 2024 or later.

\subsection{The sea that buildings hide}

On this coast, buildings hide more sea than the terrain does. In the counterfactual of Section 5, with the same viewpoints and no building in the way, 73\,\% of residential buildings would have open-view potential at their top floor, against 36\,\% with the buildings (Table 5), and 67\,\% at the ground floor, against 24\,\%. Under the main allocation, 87\,\% of dwellings would be on a floor with open-view potential and 42\,\% are: the buildings remove it for about 362,000 dwellings, more than half of those to which the terrain alone would give it (59\,\% and 21\,\% under the length-share rule; between 306,000 and 380,000 dwellings across the scenarios of Table 6). As a check on the two runs, no building sees more sea from its top floor with the buildings than without them.

\begin{figure}[!htbp]
\centering
\includegraphics[width=1.00\linewidth]{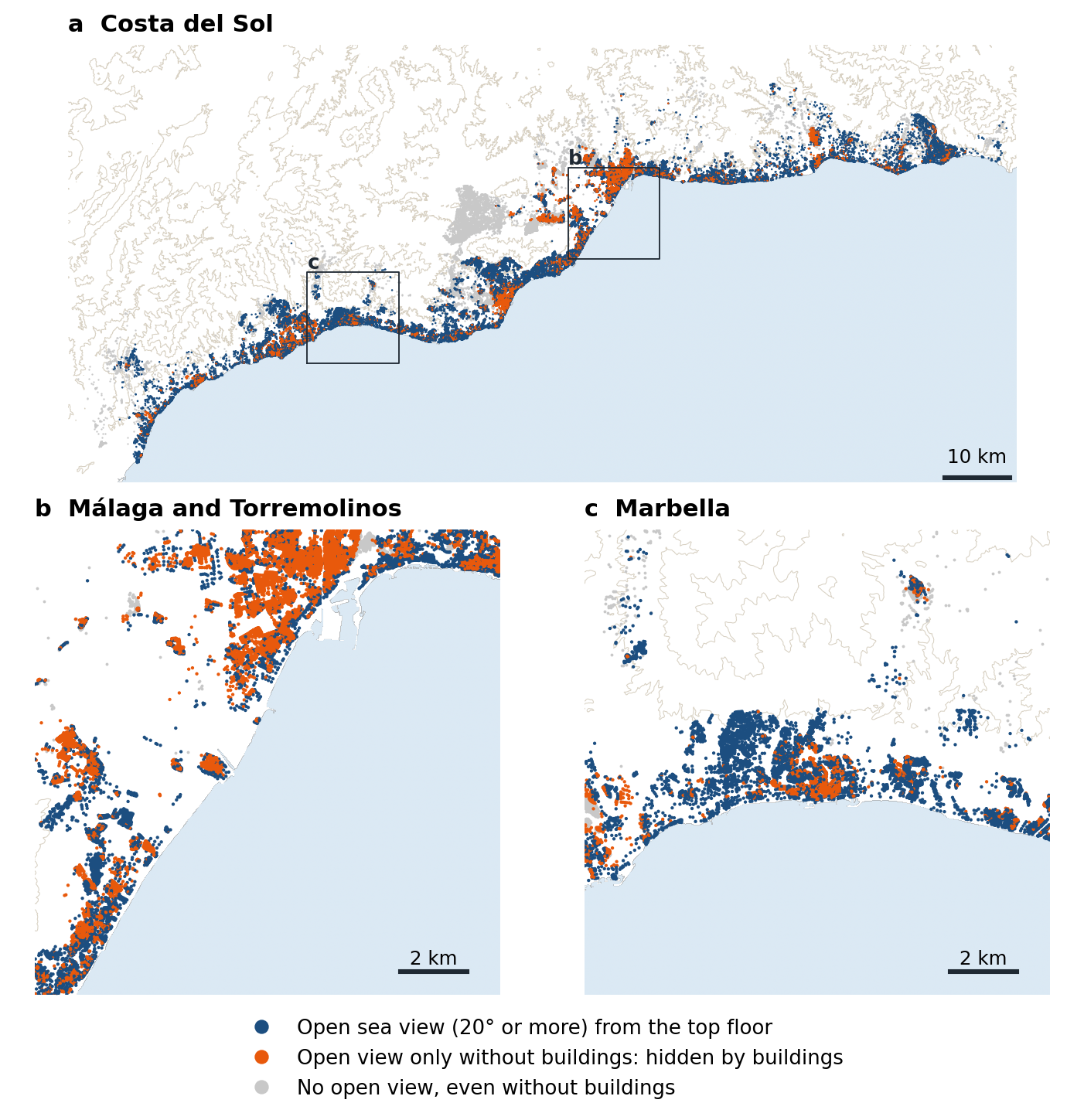}
\caption{The sea that buildings hide. Each of the 158,060 residential buildings of the Costa del Sol, from its top floor: open sea view (20\textdegree{} or more) with the buildings (blue); open view only on terrain alone, so hidden by buildings (orange); no open view even on terrain alone (grey). a: the whole coast; b: Málaga and Torremolinos; c: Marbella. Terrain and buildings, no trees.}\label{fig:6}
\end{figure}

The loss is largest where the sea is near and the town is dense (Figure 6). Within 500 m of the shore, 97\,\% of buildings would have open-view potential at their top floor on terrain alone, and 52\,\% have it. Málaga holds 46\,\% of the dwellings that lose it: 67\,\% of its buildings would have open-view potential at the top floor and 18\,\% have it. In buildings of ten floors or more, 99\,\% would have it at the top floor and 73\,\% do, while at the ground floor 91\,\% would and 14\,\% do.

Without the buildings, most of the modelled income gradient disappears. On terrain alone, the poorest fifth of dwellings would have open-view potential about as often as the richest (86\,\% against 89\,\%), also between 1 and 3 km from the shore (89\,\% against 90\,\%), where the modelled gap is widest. At section level (M2), the coefficient on log income falls from 52.5 to 3.2 (WCR \textminus{}18.6 to 34.9), and their difference, estimated directly as the coefficient of the section-level difference between the two outcomes, is 49.2 (WCR 32.2 to 71.3). With 87\,\% of dwellings open on terrain alone the linear outcome is close to saturation, so two checks matter: the fractional logit, which models the bound, gives an average marginal effect of 1.6 on terrain alone against 58.7 with buildings; and the mean degrees of sea, which do not saturate, give a terrain-only coefficient of 12.8 against 39.4, about a third of it, with a difference of 26.6 degrees (WCR 23.6 to 49.0). Beyond 3 km the terrain itself favours the richer sections (39\,\% against 69\,\% on terrain alone), where houses stand on the slopes.

This is a geometric decomposition, not a causal estimate. It holds viewpoints and heights fixed, and both are themselves products of development; it removes every building, including the other blocks of each building; and it measures what the built fabric takes as a whole, not what any neighbour, planning decision or social process takes. What it shows is that, near the sea, most of the modelled difference in view potential between richer and poorer sections lies in what stands in front of them rather than in the shape of the coast, for this model, these allocation scenarios and these outcomes.

\section{Validation}

The model is tested against four references, each for a different part of it: building heights against LiDAR (7.1); the façade-and-floor angles against a trace over a LiDAR surface (7.2); the sea seen from photographed dwellings against their photographs (7.3); and the tracer against a standard viewshed algorithm, with numerical and structural choices varied (7.4). None of them is an observation of the angle of sea that a resident sees, which this paper does not have.

\subsection{Building heights against LiDAR}

Heights from the rule of 3.0 m per floor plus 1.0 m show little median bias against airborne LiDAR, but individual parts differ by about a metre, and some by much more (Figure 7). The LiDAR height of a cadastral part is the 75th percentile of the normalised building surface inside it, shrunk by 1 m; for the 9,844 parts built in 2005 or earlier that the LiDAR detects (at least 2 m), the median difference from the rule is 0.0 m, the median absolute difference 1.0 m, and 85\,\% fall within 3 m. One-floor parts are taller in reality (median 5 m against 4 m), because of pitched roofs. Replacing the rule with LiDAR heights wherever a part built by 2014 is detected changes the class of 2.9–4.1\,\% of the pilot's buildings at every level and both thresholds. The LiDAR is dated: in each year of construction from 2010 to 2015, 53–69\,\% of parts are detected, against 2–23\,\% from 2016 on.

\begin{figure}[!htbp]
\centering
\includegraphics[width=0.85\linewidth]{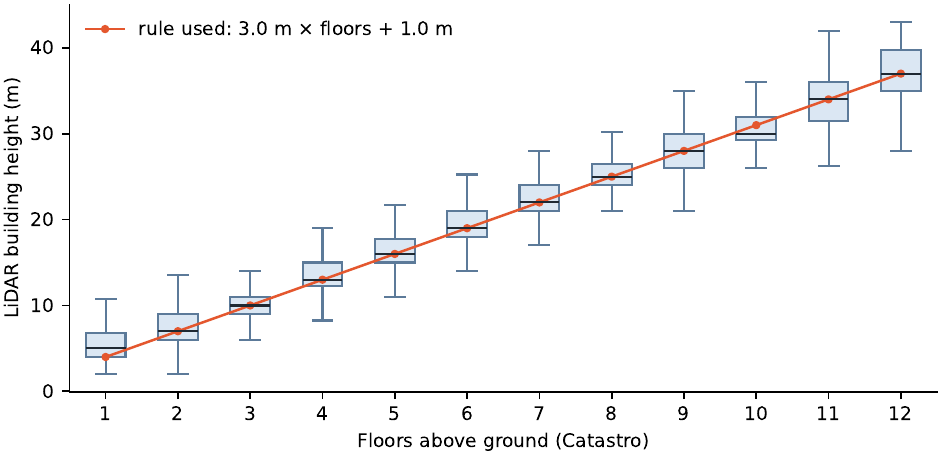}
\caption{LiDAR height of cadastral building parts by number of floors (boxes: quartiles; whiskers: 1.5 $\times$ IQR) against the rule used, 3.0 m $\times$ floors + 1.0 m. Parts built in 2005 or earlier.}\label{fig:7}
\end{figure}

\subsection{Façade-and-floor angles against a LiDAR surface}

Heights are only one of the ways in which LoD1 blocks simplify the city. To measure the angular error of the scores themselves, the pilot's 52,971 viewpoints were traced again, with the same rays, over a surface in which the cadastral blocks are replaced by the normalised building surface of the LiDAR: real roof shapes, and every structure the LiDAR classifies as a building, whether or not the cadastre records it. Inside the cadastral parts that the LiDAR does not detect, mostly built after the flight, the LoD1 block is kept. The viewpoint's own block and a 2 m ring around it stay transparent, so that registration errors do not block a façade. This surface is a reference with its own errors, a decade old and with classification noise, not an observation of the view; differences from it show how much the scores depend on how buildings are represented (Table 7).

\begin{table}[!htbp]
\centering
\footnotesize\setlength{\tabcolsep}{4pt}
\caption{Façade-and-floor scores of the model (LoD1 blocks) against the same viewpoints and rays traced over LiDAR surfaces: the pilot's 52,971 viewpoints and, at the best façade of the top floor, its 5,231 buildings. Differences are model minus reference, in degrees of sea; agreement and Cohen's \ensuremath{\kappa} for the class at 20\textdegree{}. The last row adds LiDAR trees to both, as opaque obstacles.}\label{tab:7}
\begin{tabularx}{\linewidth}{>{\hsize=1.834\hsize\RaggedRight\arraybackslash}X>{\hsize=0.917\hsize\raggedleft\arraybackslash}X>{\hsize=0.724\hsize\raggedleft\arraybackslash}X>{\hsize=0.821\hsize\raggedleft\arraybackslash}X>{\hsize=0.869\hsize\raggedleft\arraybackslash}X>{\hsize=0.772\hsize\raggedleft\arraybackslash}X>{\hsize=1.062\hsize\raggedleft\arraybackslash}X}
\toprule
\textbf{Surface traced instead of the LoD1 blocks} & \textbf{Mean sea per viewpoint, model / reference (\textdegree{})} & \textbf{Mean difference (\textdegree{})} & \textbf{90th percentile of absolute difference (\textdegree{})} & \textbf{Same class at 20\textdegree{} (\ensuremath{\kappa})} & \textbf{Top floor: mean difference (\textdegree{})} & \textbf{Buildings with open view from the top floor: model / reference} \\ \midrule
LiDAR roofs, inside the cadastral parts only & 16.5 / 16.3 & 0.2 & 8 & \mbox{96.7\,\%} (0.92) & 0.8 & \mbox{62.0\,\%} / \mbox{62.7\,\%} \\
LiDAR buildings, without patches under 50 m\textsuperscript{2} outside the parts & 16.5 / 13.6 & 2.9 & 12 & \mbox{94.0\,\%} (0.84) & 6.1 & \mbox{62.0\,\%} / \mbox{56.7\,\%} \\
LiDAR buildings, all & 16.5 / 11.9 & 4.6 & 16 & \mbox{91.7\,\%} (0.77) & 9.0 & \mbox{62.0\,\%} / \mbox{53.9\,\%} \\
LiDAR buildings and trees, against the model with opaque trees & 4.7 / 4.8 & \textminus{}0.1 & 6 & \mbox{96.2\,\%} (0.75) & 0.0 & \mbox{36.1\,\%} / \mbox{36.6\,\%} \\
\bottomrule
\end{tabularx}
\end{table}

Inside the cadastral footprints, the LoD1 blocks and the LiDAR roofs give nearly the same angles: with the LiDAR surface restricted to the cadastral parts, the mean difference is 0.2\textdegree{} per viewpoint and 0.8\textdegree{} at the best façade of the top floor, and 96.7\,\% of viewpoints fall in the same class at 20\textdegree{} (Cohen's \ensuremath{\kappa} 0.92). The difference comes from what lies outside the footprints: a third (33\,\%) of the LiDAR's building cells are more than 2 m from any cadastral building. They are walls, sheds, pergolas, unrecorded or demolished structures and classification noise, in patches with a median area of 8 m\textsuperscript{2}. With all of them, the model overstates the sea by 4.6\textdegree{} per viewpoint on average (median difference 0\textdegree{}; 90th percentile of the absolute difference 16\textdegree{}) and by 9.0\textdegree{} at the best façade of the top floor, and the share of buildings with an open view from the top floor falls from 62.0\,\% to 53.9\,\%; keeping only the patches of 50 m\textsuperscript{2} or more outside the footprints, by 6.1\textdegree{} and to 56.7\,\%. At 20\textdegree{}, 91.7\,\% of viewpoints agree (\ensuremath{\kappa} 0.77), and when the model disagrees it is mostly by finding sea that the LiDAR surface hides (0.91 specificity, 0.96 sensitivity against it). With LiDAR trees on both sides, the means match (21.5\textdegree{} and 21.5\textdegree{} at the top floor), because the trees hide much of the same sea. The angular errors of LoD1 are thus small where the cadastre maps the buildings and positive where it does not, a likely contributor to the false positives of Section 7.3.

\subsection{Sea views against listing photographs}

\textbf{Frame and matching.} Property portals usually show a home's position only as an approximate area, so the frame of this validation is the homes for sale on thinkSPAIN, a property portal, between Manilva and Nerja whose position is exact or at street level: 403 listings on 15 October 2026, 379 dwellings after removing garages, plots, premises and duplicates. Each listing was matched to a cadastral building without looking at the model: the nearest residential building within 80 m of an exact position, or 120 m of a street-level one, whose year of construction is within one year of the advertised one and whose floors fit (for a house, its floors; for a flat, at least two dwellings and enough floors for the advertised one). Where an exact position had no building with a matching year, usually a new building not yet in the cadastre, the nearest residential building within 25 m was accepted (24 cases); listings without a fitting building were dropped. Crossing a street can change a view completely, so the tolerances matter; in the main sample the median distance from the listing's position to its building is 14 m (90th percentile 56 m), and the robustness checks below restrict the sample to the most reliable matches.

\textbf{Labels.} The photographs of each listing were inspected as contact sheets and labelled before any model output was joined to them: sea visible from the dwelling; partial (a sliver or a distant sea); no sea, although the photographs show the outside; or undetermined. In this first labelling, three AI models (Claude, Anthropic; Gemini, Google; ChatGPT, OpenAI) and the author labelled the photographs, and each listing took the majority label, with ties decided by the author, who built the model. Only the final label of each listing was recorded, so agreement between those four labellers cannot be computed. The first-round labels define the sample, as in earlier versions, and are otherwise used only for comparison; the reference labels come from the blinded relabelling below. Thirty-four listings were undetermined, most of them new builds shown only in renderings, and four duplicated another dwelling, leaving 115: 53 with sea, 12 partial and 50 without. Twenty-two flats that do not state their floor are left out of the main analysis (scored from the top floor, they give an AUC of 0.76 with the rest). The main sample has 93 dwellings in 92 buildings and 12 municipalities: 46 flats and 47 houses. Three houses that only the relabelling decides are not added (with them, the AUC is 0.79 on 96 dwellings).

\textbf{Blinded relabelling.} To document the labels and test their dependence on the first labellers, the listings were labelled again under a written protocol (supplementary Section S5): the 115 labelled in the first round and 33 of the 34 it left undetermined. Five AI models of two developers did so, each run separately with the same instructions: Claude Opus 5.5, Claude Sonnet 5 and Claude Haiku 4.5 (Anthropic), and Gemini 3.1 Pro and Gemini 3.8 Flash (Google). Each saw only the listing's photographs (contact sheets and enlargements at 640 $\times$ 480 for the Anthropic models, every photograph inline for the Gemini models), under an anonymous code in random order: no listing text, position, model output or earlier label. Besides the label, each recorded the photographs that show the sea and whether the sea is seen only from a communal area. The reference label gives each developer one vote, the majority of the three Anthropic models and Gemini 3.1 Pro; Gemini 3.8 Flash is a sensitivity check. When the two votes agree, or only one of them decides, that is the label. They disagreed on 15 listings, all of the same kind, a sliver of sea that Gemini sees and the Anthropic models do not; the author decided these ties from the anonymous contact sheets, without the model's scores, and labelled 6 of them partial and 9 without sea. The four labellers that vote agree on the label of 68\,\% of the 148 listings (Fleiss' \ensuremath{\kappa} 0.76 for the four categories) and, among the 110 listings that all four decide, on 81\,\% for sea, partial included, against no sea (\ensuremath{\kappa} 0.78); Gemini 3.1 Pro and the Anthropic majority agree on 87\,\% of the 115 listings both decide (Cohen's \ensuremath{\kappa} 0.71). These are not independent human judgements, and models of two developers need not err independently, but the labels, prompts and model versions are released. The reference agrees with the first labelling on 92\,\% of the 115 listings both call determinate (Cohen's \ensuremath{\kappa} 0.84); the 9 differences are first-round negatives that the reference calls partial. In the main sample it gives 60 dwellings with sea, partial views included (35 clear, 25 partial), and 33 without, and it is the label used in the rest of this section. Against it the model's AUC is 0.80; against each labeller alone it ranges from 0.76 (Gemini 3.1 Pro, which labels more slivers as partial) to 0.84 (Gemini 3.8 Flash), against the first-round labels it is 0.81, and without the ties decided by the author, 0.81 (84 dwellings; Table 9).

\textbf{What the sample represents.} The 93 dwellings resemble the frame they come from: 50\,\% are flats against 58\,\%; 73\,\% tick the sea-view box against 63\,\%; their median size is 152 m\textsuperscript{2} against 143 m\textsuperscript{2}, and their median distance to the sea 1.4 km against 1.5 km. Homes for sale with a precise position on this coast are mostly homes with views: 82\,\% of the frame's dwellings are in a building with open-view potential from the top floor, close to the sample's 86\,\% and far above the 36\,\% of the coast's buildings. AUC does not depend on the proportion of positives as long as the scores within each class are distributed as in the target population, but selecting other kinds of homes can change those distributions \citep{leeflang2013}, so the AUC of this sample need not transport to the housing stock. Reweighting the 93 dwellings to the composition of the coast's dwellings by type and distance to the sea gives an AUC of 0.81 (95\,\% CI 0.68–0.91; effective sample 60), a sensitivity analysis on those two observed dimensions, not a regional validation.

\textbf{Label errors.} Labels made from marketing photographs can be wrong: a view may be missing from the photographs, a photographed view may come from a communal terrace, or a listing may be matched to the wrong building. Random label errors bias AUC towards 0.5: flipping 5\,\% or 10\,\% of the labels at random lowers it from 0.80 to 0.77 and 0.74 on average. Errors correlated with the score are more worrying, above all views from communal terraces or roofs, which would favour a score read at the building's best façade. The four labellers that vote flagged the sea as seen only from a communal area, or possibly so, in 5 dwellings of the main sample; leaving them out gives an AUC of 0.80.

\textbf{Scores.} For a flat advertised on floor k, every free façade of its building is read at its traced floor nearest to k + 1, and the score is the best of them; for a house, it is $S_b^{\mathrm{top}}$ of Eq. (2). Façades of lower wings that do not reach floor k + 1 are read at their own top floor. With the first-round labels, restricting the score to the façades that reach k + 1, as Eq. (3) does for dwellings, gives an AUC of 0.815 against 0.814 (paired 95\,\% CI \textminus{}0.01 to +0.02), and tracing the advertised floor exactly instead of the nearest traced floor, 0.819 (\textminus{}0.01 to +0.02).

\textbf{Yes/no answers at fixed thresholds.} The primary result is how often the model is right at the thresholds used throughout the paper (Table 8). 4\textdegree{} and 20\textdegree{} have been the paper's thresholds since its first version, not values chosen for this table, but they were not registered before the photographs were seen; 45\textdegree{} and 60\textdegree{} are exploratory. At 20\textdegree{}, sensitivity is 0.90 and specificity 0.45, a balanced accuracy of 0.68; in this sample, where 65\,\% of dwellings see the sea, 0.75 of the dwellings the model calls positive are photo-positive. These are screening results, not an inventory: the model scores the building's best façade at the dwelling's floor, not the dwelling's own windows, so a flat at the back of a sea-facing block is scored with the front, and the unmapped structures of Section 7.2 likely add further false positives. The photographs show whether the sea is visible, not how many degrees of it, so they validate no threshold as an angle.

\begin{table}[!htbp]
\centering
\footnotesize\setlength{\tabcolsep}{4pt}
\caption{The model and the sellers' checkbox as yes/no answers against the photographs (93 dwellings; 60 with sea, 33 without; reference labels). The model answers `sea' when the dwelling's score reaches the threshold; 4\textdegree{} and 20\textdegree{} are the thresholds used throughout the paper, 45\textdegree{} and 60\textdegree{} are exploratory. Counts are true positives / false negatives / false positives / true negatives. Wilson 95\,\% intervals; for balanced accuracy, 4,000 bootstrap resamples. Predictive values depend on the sample's prevalence (65\,\%) and do not transport to the housing stock.}\label{tab:8}
\begin{tabularx}{\linewidth}{>{\hsize=1.302\hsize\RaggedRight\arraybackslash}X>{\hsize=1.302\hsize\raggedleft\arraybackslash}X>{\hsize=0.921\hsize\raggedleft\arraybackslash}X>{\hsize=0.921\hsize\raggedleft\arraybackslash}X>{\hsize=0.851\hsize\raggedleft\arraybackslash}X>{\hsize=0.851\hsize\raggedleft\arraybackslash}X>{\hsize=0.851\hsize\raggedleft\arraybackslash}X}
\toprule
\textbf{Score} & \textbf{Counts} & \textbf{Sensitivity} & \textbf{Specificity} & \textbf{Balanced accuracy} & \textbf{Positive predictive value} & \textbf{Negative predictive value} \\ \midrule
Model, 4\textdegree{} or more & 59 / 1 / 28 / 5 & 0.98 (0.91–1.00) & 0.15 (0.07–0.31) & 0.57 (0.51–0.64) & 0.68 (0.57–0.77) & 0.83 (0.44–0.97) \\
Model, 20\textdegree{} or more & 54 / 6 / 18 / 15 & 0.90 (0.80–0.95) & 0.45 (0.30–0.62) & 0.68 (0.59–0.77) & 0.75 (0.64–0.84) & 0.71 (0.50–0.86) \\
Model, 45\textdegree{} or more & 43 / 17 / 5 / 28 & 0.72 (0.59–0.81) & 0.85 (0.69–0.93) & 0.78 (0.69–0.86) & 0.90 (0.78–0.95) & 0.62 (0.48–0.75) \\
Model, 60\textdegree{} or more & 39 / 21 / 4 / 29 & 0.65 (0.52–0.76) & 0.88 (0.73–0.95) & 0.76 (0.68–0.84) & 0.91 (0.78–0.96) & 0.58 (0.44–0.71) \\
Sellers' checkbox & 57 / 3 / 11 / 22 & 0.95 (0.86–0.98) & 0.67 (0.50–0.80) & 0.81 (0.72–0.89) & 0.84 (0.73–0.91) & 0.88 (0.70–0.96) \\
\bottomrule
\end{tabularx}
\end{table}

\begin{figure}[!htbp]
\centering
\includegraphics[width=0.95\linewidth]{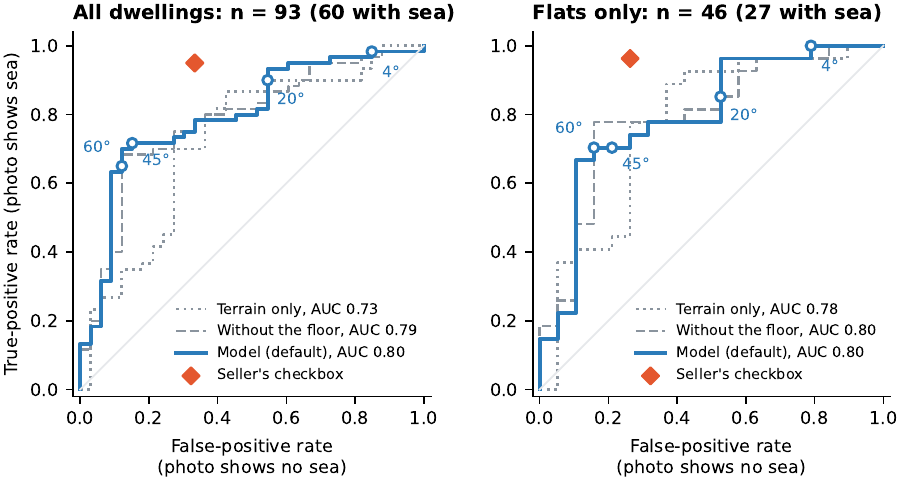}
\caption{ROC curves against the listing photographs (reference labels) for the default model and two ablations, all dwellings (93) and flats (46). Circles: the model's operating points at 4\textdegree{}, 20\textdegree{}, 45\textdegree{} and 60\textdegree{} of sea; diamond: the sellers' checkbox, a single yes/no answer.}\label{fig:8}
\end{figure}

\textbf{The sellers' checkbox.} The checkbox is a single yes/no answer, so its AUC is its balanced accuracy and the like-for-like comparison is at fixed operating points. At the checkbox's rate of false positives (11 of 33), the model's sensitivity is 0.78 against the checkbox's 0.95; at the checkbox's sensitivity, its specificity is 0.39 against 0.67. As a yes/no answer the checkbox is better, although it comes from the same seller as the photographs. The model needs no seller and covers dwellings that are not for sale.

\textbf{Ranking.} As a ranking, the model separates homes with and without a sea view better than distance to the sea, ground elevation or a rule-of-thumb heuristic (Table 9, Figure 9). Removing the buildings, so that only the terrain blocks the view, lowers the AUC from 0.80 to 0.73, and looking from the top floor of every building instead of the advertised floor lowers it to 0.79; among flats, it is 0.80 with the floor and 0.80 without. With the reference labels neither paired interval excludes zero, so on these photographs the buildings and the floor raise the ranking only within sampling error. LiDAR trees at half height raise the AUC to 0.83, and that paired interval does exclude zero; opaque trees leave it at 0.80.

\begin{table}[!htbp]
\centering
\small
\caption{Agreement with the listing photographs as a ranking: area under the ROC curve (93 dwellings; 60 with sea, 33 without; reference labels), with 95\,\% intervals and the paired difference from the model, from 4,000 bootstrap resamples of listings. The last rows score the default model against the labels of each blinded labeller, the Anthropic majority, the reference without the ties decided by the author and the first round (Section 7.3), on the dwellings each set of labels decides. The checkbox is a single yes/no answer, so its AUC is its balanced accuracy (Table 8).}\label{tab:9}
\begin{tabularx}{\linewidth}{>{\hsize=1.569\hsize\RaggedRight\arraybackslash}X>{\hsize=0.549\hsize\raggedleft\arraybackslash}X>{\hsize=0.863\hsize\RaggedRight\arraybackslash}X>{\hsize=1.020\hsize\RaggedRight\arraybackslash}X}
\toprule
\textbf{Score} & \textbf{Dwellings} & \textbf{AUC (\mbox{95\,\%} CI)} & \textbf{Model minus this score (\mbox{95\,\%} CI)} \\ \midrule
\textbf{Model: terrain + buildings + floor (default)} & 93 & \textbf{0.80} (0.70–0.89) & — \\
Model + LiDAR trees at half height & 93 & 0.83 (0.73–0.91) & \textminus{}0.057 to \textminus{}0.005 \\
Model + LiDAR trees, opaque & 93 & 0.80 (0.70–0.88) & \textminus{}0.058 to 0.076 \\
Without the floor (top floor for every dwelling) & 93 & 0.79 (0.68–0.88) & \textminus{}0.034 to 0.075 \\
Without buildings (terrain only) & 93 & 0.73 (0.61–0.84) & \textminus{}0.025 to 0.182 \\
Distance to the sea & 93 & 0.61 (0.49–0.73) & 0.062 to 0.323 \\
Ground elevation & 93 & 0.51 (0.39–0.64) & 0.128 to 0.443 \\
Rule-of-thumb heuristic (supplementary Section S3) & 85 & 0.55 (0.47–0.63) & 0.156 to 0.388 \\
Sellers' `sea view' checkbox & 93 & 0.81 (0.72–0.89) & \textminus{}0.131 to 0.113 \\
Model, against the labels of Claude Opus 5.5 (blinded) & 93 & 0.81 (0.72–0.90) & — \\
Model, against the labels of Claude Sonnet 5 (blinded) & 89 & 0.81 (0.70–0.90) & — \\
Model, against the labels of Claude Haiku 4.5 (blinded) & 87 & 0.81 (0.71–0.90) & — \\
Model, against the labels of Gemini 3.1 Pro (blinded) & 93 & 0.76 (0.64–0.87) & — \\
Model, against the labels of Gemini 3.8 Flash (blinded) & 93 & 0.84 (0.75–0.92) & — \\
Model, against the majority of the three Anthropic models & 90 & 0.82 (0.72–0.90) & — \\
Model, against the reference without the ties decided by the author & 84 & 0.81 (0.70–0.91) & — \\
Model, against the first-round labels & 93 & 0.81 (0.72–0.90) & — \\
\bottomrule
\end{tabularx}
\end{table}

\begin{figure}[!htbp]
\centering
\includegraphics[width=0.90\linewidth]{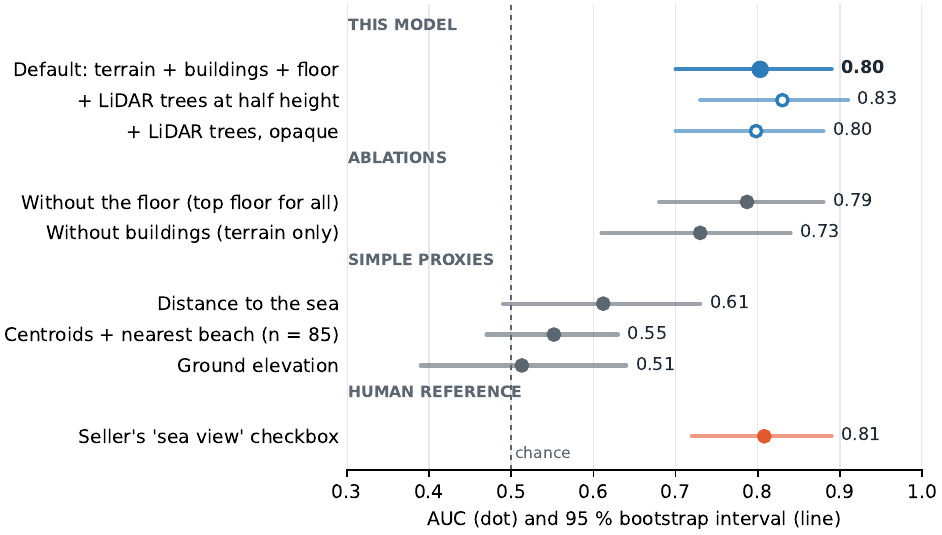}
\caption{Agreement with the listing photographs (reference labels): AUC (dot) and 95\,\% bootstrap interval (line) for the model, its tree variants, two ablations, three simple proxies, among them a rule-of-thumb heuristic, and the sellers' checkbox. 93 dwellings on the Costa del Sol; the heuristic has 85.}\label{fig:9}
\end{figure}

\textbf{Partial views.} Dwellings labelled partial score a median 50\textdegree{} (interquartile range 22–86\textdegree{}), those without sea 26\textdegree{} (8–36\textdegree{}) and those with a clear sea view 104\textdegree{} (80–148\textdegree{}). Partial views spread widely between the other two, while photo-negative dwellings spread from 0\textdegree{} (15\,\% of them) to well above the threshold (55\,\% reach 20\textdegree{}), the pattern expected from scoring the best façade. The main analysis counts partial views as sea; counting them as no sea gives an AUC of 0.84, and leaving them out 0.88; as an ordered outcome, the score's Kendall \ensuremath{\tau}-b with the three labels is 0.49.

\textbf{Robustness.} The AUC is 0.82 in Marbella (23 dwellings) and 0.80 elsewhere (70), 0.80 for flats and 0.82 for houses. Restricted to the most reliable matches, it is 0.79 for exact positions (62 dwellings), 0.80 for exact positions with a matching year (54) and 0.81 when the building is also within 30 m (37). If the 11 dwellings whose checkbox claims sea but whose photographs show none did see it, the AUC would be 0.77.

\subsection{Robustness of the geometry}

Each modelling choice was changed in turn for the pilot's buildings and carried to buildings, dwellings, the pilot's income gradient and, where re-traced, the photographs (supplementary Tables S3 and S4); the consequential ones were run for the whole coast (Table 10).

\textbf{Numerical choices.} Rays every 1\textdegree{} instead of 2\textdegree{}, a near-field step half as long and a sea mask without its 16 m edge strip change the class of at most 1.7\,\% of buildings and the pilot's share of dwellings with open-view potential by at most 0.5 points under either allocation rule. Four tiles (28,720 buildings) traced with buildings blocking up to 4 km instead of 2 km change class for 0.1–0.7\,\% of buildings. For 60 random viewpoints of the pilot, the radial sweep of GRASS r.viewshed \citep{haverkort2009}, as implemented in xarray-spatial 0.5.3, run on the same 2 m surface agrees with the model on whether any sea within 2 km is visible on 99.8\,\% of 5,160 rays (Cohen's \ensuremath{\kappa} 0.96); this verifies the line-of-sight implementation, not the surface.

\textbf{Structural assumptions.} These change what the surface represents, not how precisely it is computed, and they matter more.

\begin{itemize}

\item \emph{Field of view and viewpoint.} Keeping the rays within $\pm$ 60\textdegree{} or $\pm$ 45\textdegree{} of the normal changes the class of 6.1\,\% and 10.2\,\% of the pilot's buildings at the top floor. Placing the viewpoint on the plane of the wall, with its own block opaque beyond 1 m, moves the pilot's dwelling share from 50.4\,\% to 48.7\,\%.

\item \emph{Floors of a façade.} A radius of 1.5 m instead of 6 m changes the class of 9.7\,\% of buildings at the top floor, mostly stepped buildings whose top floor drops to that of a lower part in front. In the 378 stepped buildings of the pilot built by 2005 whose highest and lowest parts the LiDAR sees, the step goes the same way in 92\,\%.

\item \emph{Own block, viewpoints and roofs.} The own block opaque beyond 4 m, three viewpoints along each octant's longest edge and flat roofs each change the class of at most 3.6\,\% of buildings.

\item \emph{Vegetation.} Opaque LiDAR trees lower the pilot's dwelling share from 50.4\,\% to 33.5\,\% and its income gradient from 120 (SE 28) to 32 (SE 20), because its richer sections are hillside houses among gardens; trees at half their height, to 44.7\,\% and 85 (25). Neither is a calibrated porosity or a bound.

\item \emph{Distance.} Weighting each sea ray by min(1, 1 km / d) narrows the gap between the richest and poorest fifths from 15.6 to 7.3 points, because the open views of the richest look at sea a median 1.1 km away, against 0.7 km for the poorest.

\end{itemize}

Against the first-round labels of the photographs, the variants give AUCs from 0.78 to 0.83, against 0.81 for the default (supplementary Table S4); the photographs cannot choose among most of them (with the reference labels, trees at half height rank slightly better; Section 7.3), and the defaults were not chosen for fit.

\textbf{Combinations, for the whole coast.} Table 10 crosses the geometric and vegetation scenarios run for the whole coast with the two allocation rules, for the two headline results. The share of dwellings with open-view potential ranges from 26.9\,\% to 42.4\,\% with the best façade and from 12.4\,\% to 21.2\,\% with the length share; the combination of conservative geometry and opaque trees, with the length share, gives 12.4\,\%. These are scenarios, not a confidence interval: they change what the model represents. Table A4 adds any sea and the mean degrees of sea.

\begin{table}[!htbp]
\centering
\footnotesize\setlength{\tabcolsep}{4pt}
\caption{The two headline results under each geometric and vegetation scenario run for the whole coast, crossed with the two allocation rules (nearest traced floor, 20\textdegree{} or more): share of the coast's dwellings on a floor with open-view potential, and income coefficient with municipality fixed effects (M2; points per unit of log income) with its WCR 95\,\% interval in brackets. Scenarios change what the model represents; their range is not a confidence interval. The last rows give other statistical intervals for the default with the best façade, and the terrain-only decomposition.}\label{tab:10}
\begin{tabularx}{\linewidth}{>{\hsize=1.846\hsize\RaggedRight\arraybackslash}X>{\hsize=0.615\hsize\raggedleft\arraybackslash}X>{\hsize=0.962\hsize\raggedleft\arraybackslash}X>{\hsize=0.615\hsize\raggedleft\arraybackslash}X>{\hsize=0.962\hsize\raggedleft\arraybackslash}X}
\toprule
\textbf{Scenario} & \textbf{Dwellings, best façade} & \textbf{Income coefficient, best façade [WCR]} & \textbf{Dwellings, length share} & \textbf{Income coefficient, length share [WCR]} \\ \midrule
Default (Table 3) & \mbox{42.4\,\%} & 52.5 [14.4, 92.9] & \mbox{21.2\,\%} & 23.2 [8.6, 38.2] \\
Floors of a façade within 1.5 m & \mbox{39.1\,\%} & 50.6 [11.0, 91.6] & \mbox{19.3\,\%} & 21.8 [9.9, 34.5] \\
LiDAR trees at half height & \mbox{38.7\,\%} & 46.1 [11.0, 86.3] & \mbox{18.5\,\%} & 18.1 [7.6, 29.6] \\
LiDAR trees, opaque & \mbox{31.4\,\%} & 33.5 [9.9, 68.8] & \mbox{14.1\,\%} & 10.7 [4.5, 16.7] \\
Field of view $\pm$ 60\textdegree{} & \mbox{39.1\,\%} & 49.3 [6.7, 94.3] & \mbox{15.4\,\%} & 16.2 [4.1, 28.0] \\
Field of view $\pm$ 45\textdegree{} & \mbox{36.7\,\%} & 47.7 [3.5, 93.7] & \mbox{12.5\,\%} & 13.2 [4.1, 22.9] \\
Sea extent weighted by distance (relative to 1 km) & \mbox{26.9\,\%} & 28.0 [\textminus{}8.1, 64.8] & \mbox{12.5\,\%} & 10.5 [\textminus{}0.2, 22.0] \\
Combined: 1.5 m radius, own block opaque beyond 4 m, flat roofs, trees at half height & \mbox{36.5\,\%} & 44.9 [4.4, 88.4] & \mbox{16.5\,\%} & 16.4 [8.3, 26.0] \\
The same combination with opaque trees & \mbox{29.6\,\%} & 32.9 [3.7, 72.0] & \mbox{12.4\,\%} & 9.7 [3.2, 15.5] \\
\emph{Default, best façade: CV3 with t(20)} & — & 9.1 to 95.8 & — & — \\
\emph{Default, best façade: Conley standard errors at 2, 5 and 10 km} & — & 7.5, 9.3, 8.7 & — & — \\
\emph{Default, best façade: 5 km spatial blocks, four grid origins} & — & 35.2 to 82.5 (widest) & — & — \\
\emph{Terrain only (a decomposition, not an uncertainty)} & \mbox{87.2\,\%} & 3.2 [\textminus{}18.6, 34.9] & \mbox{59.4\,\%} & 0.3 [\textminus{}3.2, 16.4] \\
\bottomrule
\end{tabularx}
\end{table}

\section{Discussion}

\subsection{What the model measures, and how well}

The model measures the sea-view potential of a façade at a floor. That quantity is well supported where the cadastre maps the buildings: LoD1 heights show no median bias against LiDAR, angles traced over LiDAR roofs inside the footprints agree closely with the model's, and the tracer agrees with a standard viewshed algorithm. It is biased upwards where structures the cadastre does not record stand in the way, by 6.1–9.0\textdegree{} of sea at the top floor of the pilot's buildings on average, and trees, left out of the default, are the largest single uncertainty.

Against photographs, the model ranks homes with and without a sea view well (AUC 0.80 against labels from AI models of two developers, with ties decided by the author, and 0.81 against the first-round labels), much better than distance to the sea or ground elevation; the buildings and the floor raise the ranking, though with these labels within sampling error. As a yes/no answer at 20\textdegree{} it is a sensitive screen with poor specificity (balanced accuracy 0.68), less specific than the sellers' checkbox, because the best façade of a building at a floor is not the view of every dwelling on it. The photographs validate neither the 20\textdegree{} threshold nor the share of dwellings that see the sea.

Building and dwelling figures are aggregations. The building figures say whether some dwelling at a level could see the sea; the dwelling figures depend on how dwellings are allocated to façades and floors, and range from 12\,\% to 45\,\% of the coast's dwellings under the scenarios tested (Tables 6 and 10). They are model outputs for comparison between areas and scenarios, not an inventory of who sees the sea.

As an application, higher-income census sections have more modelled view potential within municipalities, and on terrain alone that difference largely disappears. With an effective number of about two clusters, the size and precision of the association are uncertain, and they depend on representation: it is a descriptive, conditional association, and the decomposition is geometric, not an account of how planning or construction produced it.

\subsection{Uses}

The output is a set of attributes per building, façade and floor (sea extent, bearing and distance; mountain extent and dominant summit; openness; view façade), joinable to any dataset with a cadastral reference. Hedonic studies can include floor-level view variables for every transaction instead of sellers' declarations \citep{benson1998,bourassa2004}, and separate what is visible from what is near \citep{walls2015}; because the scores rank better than they classify, they suit such continuous or ranked uses better than yes/no flags. Health studies can move from proximity to visibility of blue space at the home for whole cohorts \citep{nutsford2016,white2021}. Planning can screen the façades where the view out matters most and measure the views a proposed building would take from those behind it by running the model with and without it; this is not a test of EN 17037 \citep{cen2018}. If disclosed information changes prices \citep{pope2008}, comparable view data for every dwelling, not only those advertised, is a public-interest product.

\subsection{Limitations}

\begin{itemize}

\item \textbf{The dwelling's own view.} At each floor the model reports the building's façades, not each dwelling's windows; dwellings are allocated by scenario, not observed. Joining the cadastre's property records by floor (Section 4.4) would remove the even spread over floors, but not the assignment of dwellings to façades.

\item \textbf{The reference for sea views.} The photographed dwellings are listings with precise positions on one coast, not a random sample of homes, and their AUC need not transport to the stock. Marketing photographs can miss a view or show one from a communal terrace, and matching tolerances of 80–120 m can attach a listing to the wrong building. All labels, first and blinded, come from AI models and, in the first round and in the 15 ties of the reference, the author, who built the model and runs a website that displays its scores (Competing interests). An independent sample with verified building, floor and viewing direction, labelled by people blind to the model with settings frozen beforehand, is the validation this paper lacks: to estimate an AUC near 0.80 within $\pm$ 0.05 it would need about 290 dwellings, and calibrated panoramas or surveyed sightlines would test the angles.

\item \textbf{Unmapped structures and trees.} The cadastre misses structures that block views, and the default leaves trees out; both bias the model towards seeing more sea (Section 7.2), and the vegetation layer is a decade old. A recent LiDAR flight, processed for the validation area, would allow a calibrated porosity per vegetation type \citep{murgoitio2014}.

\item \textbf{Geometry.} Only three floors are traced; one viewpoint per octant simplifies long façades; balconies, setbacks and windows are not modelled; and the field of view of $\pm$ 86\textdegree{} is that of an observer at a balcony rail.

\item \textbf{Degrees of sea.} The main extent counts a distant sliver like near sea; the distance-weighted extent and the released distance to the first visible sea address this. Haze and season are not modelled, and the mountain and openness metrics are not validated.

\item \textbf{Income.} The income analysis is an area-level application with resident incomes from tax records, a stock that includes second homes, and few effective clusters (Section 6.4).

\end{itemize}

\subsection{Applying the method elsewhere}

The Spanish cadastre records floors for every building part in the territory it covers, all of Spain except the Basque Country and Navarre, whose own cadastres publish equivalent data, and the IGN terrain models cover the whole country. At 9.6 core-hours per million buildings, a stock of ten million buildings would take about 100 core-hours in independent tiles. That the model runs is not that it is equally accurate: it has been tested on one coast of tall blocks and hillside houses, and dense historic centres, flat coasts and other building types need their own checks. Elsewhere, the method applies where footprints come with floor counts, which is the case in some national cadastres and in parts of harmonised collections such as EUBUCCO \citep{milojevicdupont2023}, and not everywhere; where floors are missing, global LoD1 heights, with an error of about 4 m in Europe \citep{zhu2025}, could stand in for them, untested here.

\section{Conclusions}

Sea-view potential by façade and floor can be computed for a whole region from two open datasets: cadastral footprints with floor counts and a national terrain model. Where the cadastre maps the buildings, its angles agree closely with a trace over LiDAR roofs; where it does not, and wherever trees stand, it overstates the sea. Against listing photographs labelled by AI models of two developers, with ties decided by the author, it ranks homes with and without a sea view with an AUC of 0.80, but as a yes/no answer it is a sensitive screen with poor specificity. Dwelling figures are scenarios, from 12\,\% to 45\,\% of the coast's dwellings with open-view potential, not measurements. As an application, higher-income census sections tend to have more modelled view potential within municipalities, with uncertain size and fragile inference, and on terrain alone the difference largely disappears. Linking dwellings to their floors with the cadastre's property records, a recent vegetation layer and an independent validation sample with verified floors and viewing directions are the next steps. Released by façade and floor, with their assumptions stated, these scores can become a reusable variable for research on housing, health and the built environment.

\section*{Data and code availability}

The reviewer archive submitted with this paper (who-sees-the-sea-reviewer-archive.zip; SHA-256 in the submission letter) contains the whole release, and its README gives one command for each check. \textbf{Code:} the model; the scripts that reproduce every table and figure from the public inputs; a test area, shipped as plain GeoJSON, with its expected output; and the computational environment, with pinned versions and a lock file, under the MIT licence. \textbf{Data:} under CC BY 4.0, the view attributes of the 158,060 residential buildings per building and level; the degrees of sea of each of the 1,349,970 façade viewpoints and traced floors, with buildings, on terrain alone and with trees, with the 14-character cadastral reference, octant, free length and floor count of each façade (Section 4.4); the dwelling shares under every allocation scenario; the section-level table behind Tables 6 and 10 and Appendix A; and the LiDAR comparison of Section 7.2. \textbf{Validation:} for each of the 148 relabelled listings, under an anonymous code, the municipality, dwelling type, advertised floor, position precision, matching rule and distance, the first label, the five blinded labels (four of them with communal-view flags), the author's decision on ties, the reference label, and the model scores, which reproduce Tables 8 and 9 except the sellers' checkbox; the labelling protocol and prompt. The table carries no listing or building identifier, but its scores could be joined to the façade table to recover the buildings, so it is in the reviewer archive and is shared with editors and reviewers only, as are the listing identifiers, building matches and checkbox values, for audit; the public deposit carries the aggregated validation results. No listing photograph or text is redistributed. The inputs are open data of the Dirección General del Catastro, the Instituto Geográfico Nacional and the Instituto Nacional de Estadística, downloadable with the scripts; the archive lists the version, download date and SHA-256 checksum of every input file. A script reproduces the dwelling shares and the regressions M1 and M2 with their WCR intervals from the released tables alone, and the AUCs from the reviewer archive, in minutes on a laptop. The public release is deposited in Zenodo (\url{https://doi.org/10.5281/zenodo.24106387}) and linked from \url{https://casas.es/en/research/who-sees-the-sea}. The \href{https://casas-es.github.io/datos/malaga/calles-de-la-costa/}{streets of the coast}, each linked to its page at casas.es, are listed on the CASAS.ES open-data site.

\section*{Competing interests}

The author runs CASAS.ES, a free website with data on every building in Spain, which displays view scores derived from this model. No external funding was received.

\section*{Acknowledgements}

The model uses data from the Dirección General del Catastro and the Instituto Geográfico Nacional (\textcopyright{} IGN, licence compatible with CC BY 4.0); the analyses use the census sections and household income atlas of the Instituto Nacional de Estadística, and energy certificates from the Registro de Certificados Energéticos de Andalucía. Listing data were consulted on thinkSPAIN (thinkspain.com).

\section*{Declaration of generative AI use}

The code, the analyses and the drafts of this paper were prepared with an AI assistant (Claude, Anthropic), and several AI models, among them Gemini (Google) and ChatGPT (OpenAI), were used to review the text and the code. Claude, Gemini and ChatGPT labelled the listing photographs of Section 7.3 in the first round, together with the author, who decided ties; Claude Opus 5.5, Claude Sonnet 5, Claude Haiku 4.5, Gemini 3.1 Pro and Gemini 3.8 Flash labelled them again under the blinded protocol of supplementary Section S5, and the author decided the 15 listings on which the Anthropic models and Gemini 3.1 Pro disagreed. All labels were made before any model output was joined to them. The author directed the work and is responsible for the content.

\appendix

\section{Complete regression results}
\setcounter{table}{0}\renewcommand{\thetable}{\Alph{section}\arabic{table}}

Table A1 reports the estimates and inference of every model of Section 6.4 for the main outcome, Table A2 their other coefficients, samples and fit, Table A3 model M2 for other outcomes with and without buildings and their paired difference, and Table A4 the scenarios of Table 10 for any sea and the mean degrees of sea. Each section is assigned to the municipality that holds most of its dwellings. Bootstrap intervals can be asymmetric around the coefficient. With so few effective clusters, the WCR p-value need not be monotone in the hypothesised value: in one cell (terrain only, any sea, best façade; marked † in Tables A3 and A4), zero is rejected but values on both sides of it are not, so the set of values not rejected is not a single interval. M1 gives the association across the whole coast; M2 to M4 and M7 to M8 within municipalities, with different adjustments; M5 and M6 the association that remains once building age, type, height and ground elevation, partly channels, are held fixed. They are not bounds on one another.

\begin{table}[!htbp]
\centering
\scriptsize
\caption{Section-level regressions of the share of dwellings with open-view potential (best façade, nearest traced floor, 20\textdegree{}; percent) on log section income: estimates and inference. Weighted least squares, one row per census section with at least 20 dwellings, weights = dwellings; FE: municipality fixed effects. Standard errors: HC1, clustered by municipality (CV1), leave-one-municipality-out (CV3) and Conley (Bartlett kernel, 5 km). 95\,\% intervals: restricted wild cluster bootstrap-t (WCR; 9,999 replications, Webb weights, interval by test inversion) with its p-value; unrestricted (WCU); CV3 with t(G \textminus{} 1); and, for M2, square spatial blocks (499 draws, of which those where a fixed effect cannot be estimated are skipped; grid at the origin). G*: effective number of clusters \citep{carter2017}; partial leverage: share of the leverage for log income held by Málaga. Specifications and fit in Table A2.}\label{tab:A1}
\begin{tabularx}{\linewidth}{>{\hsize=2.379\hsize\RaggedRight\arraybackslash}X>{\hsize=0.828\hsize\raggedleft\arraybackslash}X>{\hsize=0.828\hsize\raggedleft\arraybackslash}X>{\hsize=0.828\hsize\raggedleft\arraybackslash}X>{\hsize=0.828\hsize\raggedleft\arraybackslash}X>{\hsize=0.828\hsize\raggedleft\arraybackslash}X>{\hsize=0.828\hsize\raggedleft\arraybackslash}X>{\hsize=0.828\hsize\raggedleft\arraybackslash}X>{\hsize=0.828\hsize\raggedleft\arraybackslash}X}
\toprule
\textbf{} & \textbf{M1} & \textbf{M2} & \textbf{M3} & \textbf{M4} & \textbf{M5} & \textbf{M6} & \textbf{M7} & \textbf{M8} \\ \midrule
Specification & Base & FE & FE + distance bands & FE + non-main & FE + controls & FE + controls + non-main & FE, without Málaga & FE, 1–3 km only \\
\textbf{Log income} & \textbf{30.2} & \textbf{52.5} & \textbf{53.9} & \textbf{47.6} & \textbf{16.5} & \textbf{17.4} & \textbf{73.5} & \textbf{53.1} \\
SE, HC1 & 5.2 & 4.4 & 4.4 & 4.2 & 4.6 & 4.4 & 9.2 & 7.3 \\
SE, clustered (CV1) & 10.4 & 7.4 & 7.0 & 6.6 & 4.6 & 4.5 & 9.7 & 15.1 \\
SE, CV3 & 33.7 & 20.8 & 14.9 & 17.6 & 4.5 & 10.6 & 10.2 & 60.0 \\
SE, Conley (5 km) & 10.6 & 9.3 & — & — & 5.1 & — & — & — \\
\mbox{95\,\%} CI, WCR & \textminus{}1.8 to 85.4 & 14.4 to 92.9 & 19.4 to 86.6 & 14.7 to 87.7 & \textminus{}6.2 to 37.0 & \textminus{}6.7 to 42.8 & 55.0 to 106.4 & \textminus{}140.6 to 147.4 \\
p-value, WCR & 0.054 & 0.026 & 0.021 & 0.027 & 0.098 & 0.073 & 0.002 & 0.172 \\
\mbox{95\,\%} CI, WCU & 8.8 to 51.7 & 26.4 to 78.5 & — & — & 10.1 to 23.0 & — & — & — \\
\mbox{95\,\%} CI, CV3 with t(G \textminus{} 1) & \textminus{}40.0 to 100.5 & 9.1 to 95.8 & 22.7 to 85.0 & 10.8 to 84.4 & 7.1 to 26.0 & \textminus{}4.7 to 39.6 & 52.0 to 94.9 & \textminus{}77.5 to 183.8 \\
\mbox{95\,\%} CI, 2.5 km spatial blocks & — & 37.3 to 67.7 & — & — & — & — & — & — \\
\mbox{95\,\%} CI, 5 km spatial blocks & — & 37.1 to 70.4 & — & — & — & — & — & — \\
\mbox{95\,\%} CI, 10 km spatial blocks & — & 40.3 to 91.2 & — & — & — & — & — & — \\
Effective clusters G* & 2.6 & 2.2 & 2.3 & 2.2 & 2.3 & 2.2 & 7.0 & 1.6 \\
Partial leverage of Málaga & 0.60 & 0.66 & — & — & 0.65 & — & — & — \\
Moran's I of residuals (sections within 2 km) & 0.85 & 0.27 & — & — & 0.18 & — & — & — \\
\bottomrule
\end{tabularx}
\end{table}

\begin{table}[!htbp]
\centering
\scriptsize
\caption{The regressions of Table A1: other coefficients, samples and fit. Controls: mean year of construction (per year), share of dwellings in single-family houses (percent), mean floors, mean ground elevation (per 100 m); non-main dwellings: percent of dwellings that are not main residences in the 2021 census. Distance bands: < 250 m (reference), 250–500 m, 0.5–1 km, 1–2 km, 2–3 km, 3–5 km, > 5 km.}\label{tab:A2}
\begin{tabularx}{\linewidth}{>{\hsize=2.379\hsize\RaggedRight\arraybackslash}X>{\hsize=0.828\hsize\raggedleft\arraybackslash}X>{\hsize=0.828\hsize\raggedleft\arraybackslash}X>{\hsize=0.828\hsize\raggedleft\arraybackslash}X>{\hsize=0.828\hsize\raggedleft\arraybackslash}X>{\hsize=0.828\hsize\raggedleft\arraybackslash}X>{\hsize=0.828\hsize\raggedleft\arraybackslash}X>{\hsize=0.828\hsize\raggedleft\arraybackslash}X>{\hsize=0.828\hsize\raggedleft\arraybackslash}X}
\toprule
\textbf{} & \textbf{M1} & \textbf{M2} & \textbf{M3} & \textbf{M4} & \textbf{M5} & \textbf{M6} & \textbf{M7} & \textbf{M8} \\ \midrule
Specification & Base & FE & FE + distance bands & FE + non-main & FE + controls & FE + controls + non-main & FE, without Málaga & FE, 1–3 km only \\
Log distance to the sea & \textminus{}11.6 & \textminus{}6.3 & bands & \textminus{}1.6 & \textminus{}17.7 & \textminus{}11.3 & \textminus{}7.8 & 15.3 \\
Year built (per year) &  &  &  &  & 0.98 & 0.88 &  &  \\
\% dwellings in houses &  &  &  &  & 0.07 & 0.01 &  &  \\
Mean floors &  &  &  &  & 0.46 & 0.86 &  &  \\
Mean ground elevation (per 100 m) &  &  &  &  & 22.62 & 20.45 &  &  \\
\% non-main dwellings &  &  &  & 0.62 &  & 0.56 &  &  \\
Municipality fixed effects & no & yes & yes & yes & yes & yes & yes & yes \\
Sections & 860 & 860 & 860 & 839 & 860 & 839 & 425 & 354 \\
Municipalities & 21 & 21 & 21 & 21 & 21 & 21 & 20 & 13 \\
R\textsuperscript{2} & 0.18 & 0.53 & 0.53 & 0.62 & 0.64 & 0.71 & 0.41 & 0.62 \\
Within-municipality R\textsuperscript{2} & — & 0.21 & 0.22 & 0.38 & 0.39 & 0.52 & 0.18 & 0.21 \\
\bottomrule
\end{tabularx}
\end{table}

\begin{table}[!htbp]
\centering
\footnotesize\setlength{\tabcolsep}{4pt}
\caption{The income coefficient for other outcomes, with buildings and on terrain alone, and their paired difference (model M2). Points per unit of log income, or degrees for the mean sea extent. In brackets, WCR 95\,\% intervals (9,999 replications); the difference is the coefficient of the section-level difference between the two outcomes, with the same inference. †: the WCR p-value is not monotone in the hypothesised value here: zero is rejected (p = 0.017) but values from \textminus{}8.7 to about \textminus{}1.3 are not, so the set of values not rejected is two intervals (supplementary S6).}\label{tab:A3}
\begin{tabularx}{\linewidth}{>{\hsize=1.690\hsize\RaggedRight\arraybackslash}X>{\hsize=0.634\hsize\raggedleft\arraybackslash}X>{\hsize=0.972\hsize\raggedleft\arraybackslash}X>{\hsize=0.676\hsize\raggedleft\arraybackslash}X>{\hsize=0.972\hsize\raggedleft\arraybackslash}X>{\hsize=1.056\hsize\raggedleft\arraybackslash}X}
\toprule
\textbf{Outcome (nearest traced floor)} & \textbf{Mean, model} & \textbf{Coefficient, model [WCR]} & \textbf{Mean, terrain only} & \textbf{Coefficient, terrain only [WCR]} & \textbf{Model minus terrain only [WCR]} \\ \midrule
Any sea (4\textdegree{} or more), best façade & \mbox{58.3\,\%} & 62.5 [45.3, 92.5] & \mbox{92.7\,\%} & 4.2 [\textminus{}8.7, 20.3]† & 58.3 [47.2, 73.0] \\
Open view (20\textdegree{} or more), best façade & \mbox{42.4\,\%} & 52.5 [14.4, 92.9] & \mbox{87.2\,\%} & 3.2 [\textminus{}18.6, 34.9] & 49.2 [32.2, 71.3] \\
Wide view (45\textdegree{} or more), best façade & \mbox{29.3\,\%} & 37.1 [\textminus{}4.6, 76.2] & \mbox{77.2\,\%} & 12.2 [\textminus{}13.2, 45.9] & 24.8 [7.3, 44.8] \\
Mean degrees of sea, best façade & 34.4\textdegree{} & 39.4 [\textminus{}2.8, 82.8] & 94.9\textdegree{} & 12.8 [\textminus{}14.9, 43.0] & 26.6 [23.6, 49.0] \\
Any sea (4\textdegree{} or more), length share & \mbox{34.0\,\%} & 34.3 [20.3, 51.4] & \mbox{69.6\,\%} & 0.6 [\textminus{}6.3, 9.3] & 33.7 [20.9, 47.2] \\
Open view (20\textdegree{} or more), length share & \mbox{21.2\,\%} & 23.2 [8.6, 38.2] & \mbox{59.4\,\%} & 0.3 [\textminus{}3.2, 16.4] & 22.9 [13.9, 31.8] \\
Wide view (45\textdegree{} or more), length share & \mbox{11.8\,\%} & 12.4 [2.4, 22.5] & \mbox{46.4\,\%} & 4.3 [\textminus{}1.9, 13.6] & 8.1 [4.3, 13.5] \\
Mean degrees of sea, length share & 14.1\textdegree{} & 13.9 [5.8, 23.0] & 48.6\textdegree{} & 0.4 [\textminus{}7.0, 6.4] & 13.4 [12.0, 24.6] \\
\bottomrule
\end{tabularx}
\end{table}

\begin{table}[!htbp]
\centering
\scriptsize
\caption{The scenarios of Table 10 for any sea (4\textdegree{} or more) and for the mean degrees of sea: share of dwellings (or mean degrees) and M2 income coefficient with its WCR 95\,\% interval, under the two allocation rules. †: the WCR p-value is not monotone in the hypothesised value here: zero is rejected (p = 0.017) but values from \textminus{}8.7 to about \textminus{}1.3 are not, so the set of values not rejected is two intervals (supplementary S6).}\label{tab:A4}
\begin{tabularx}{\linewidth}{>{\hsize=1.921\hsize\RaggedRight\arraybackslash}X>{\hsize=0.708\hsize\raggedleft\arraybackslash}X>{\hsize=1.062\hsize\raggedleft\arraybackslash}X>{\hsize=0.708\hsize\raggedleft\arraybackslash}X>{\hsize=1.062\hsize\raggedleft\arraybackslash}X>{\hsize=0.708\hsize\raggedleft\arraybackslash}X>{\hsize=1.062\hsize\raggedleft\arraybackslash}X>{\hsize=0.708\hsize\raggedleft\arraybackslash}X>{\hsize=1.062\hsize\raggedleft\arraybackslash}X}
\toprule
\textbf{Scenario} & \textbf{Any sea, best façade} & \textbf{Coef. [WCR]} & \textbf{Any sea, length share} & \textbf{Coef. [WCR]} & \textbf{Mean degrees, best façade} & \textbf{Coef. [WCR]} & \textbf{Mean degrees, length share} & \textbf{Coef. [WCR]} \\ \midrule
Default (Table 3) & \mbox{58.3\,\%} & 62.5 [45.3, 92.5] & \mbox{34.0\,\%} & 34.3 [20.3, 51.4] & 34.4\textdegree{} & 39.4 [\textminus{}2.8, 82.8] & 14.1\textdegree{} & 13.9 [5.8, 23.0] \\
Floors of a façade within 1.5 m & \mbox{54.9\,\%} & 61.4 [40.1, 93.5] & \mbox{31.8\,\%} & 33.7 [19.5, 49.9] & 30.5\textdegree{} & 35.0 [\textminus{}0.3, 73.4] & 12.7\textdegree{} & 12.6 [6.7, 19.1] \\
LiDAR trees at half height & \mbox{53.8\,\%} & 54.1 [32.1, 90.7] & \mbox{29.7\,\%} & 26.0 [16.0, 39.2] & 30.4\textdegree{} & 33.0 [\textminus{}2.3, 71.7] & 12.2\textdegree{} & 10.6 [5.2, 17.2] \\
LiDAR trees, opaque & \mbox{46.0\,\%} & 42.9 [23.4, 78.5] & \mbox{23.4\,\%} & 16.3 [6.5, 24.9] & 23.5\textdegree{} & 22.7 [3.8, 51.9] & 9.2\textdegree{} & 6.1 [1.5, 8.9] \\
Field of view $\pm$ 60\textdegree{} & \mbox{53.4\,\%} & 60.3 [35.7, 97.6] & \mbox{24.9\,\%} & 24.8 [10.2, 40.9] & 27.9\textdegree{} & 33.6 [\textminus{}4.4, 72.1] & 9.5\textdegree{} & 9.3 [3.2, 16.2] \\
Field of view $\pm$ 45\textdegree{} & \mbox{51.3\,\%} & 59.1 [32.9, 97.9] & \mbox{20.8\,\%} & 20.9 [8.7, 33.9] & 23.0\textdegree{} & 28.4 [\textminus{}2.7, 60.3] & 7.1\textdegree{} & 7.0 [2.7, 12.0] \\
Sea extent weighted by distance (relative to 1 km) & \mbox{46.2\,\%} & 54.9 [16.6, 93.8] & \mbox{24.9\,\%} & 25.7 [9.2, 43.3] & 21.7\textdegree{} & 18.6 [0.9, 41.2] & 8.9\textdegree{} & 6.5 [\textminus{}0.3, 12.3] \\
Combined: 1.5 m radius, own block opaque beyond 4 m, flat roofs, trees at half height & \mbox{50.6\,\%} & 54.1 [22.8, 92.7] & \mbox{26.6\,\%} & 23.6 [13.9, 34.5] & 28.0\textdegree{} & 32.3 [\textminus{}7.1, 73.4] & 10.8\textdegree{} & 9.9 [5.1, 15.4] \\
The same combination with opaque trees & \mbox{43.1\,\%} & 43.1 [15.4, 82.9] & \mbox{20.8\,\%} & 14.7 [4.5, 22.0] & 21.6\textdegree{} & 22.2 [\textminus{}2.2, 54.6] & 8.0\textdegree{} & 5.5 [0.1, 8.9] \\
Terrain only (decomposition) & \mbox{92.7\,\%} & 4.2 [\textminus{}8.7, 20.3]† & \mbox{69.6\,\%} & 0.6 [\textminus{}6.3, 9.3] & 94.9\textdegree{} & 12.8 [\textminus{}14.9, 43.0] & 48.6\textdegree{} & 0.4 [\textminus{}7.0, 6.4] \\
\bottomrule
\end{tabularx}
\end{table}

\FloatBarrier

\end{document}